\documentclass[11pt,a4paper]{article}

\usepackage[a4paper,text={150mm,240mm},centering,headsep=10mm,footskip=15mm]{geometry}
\usepackage[T1]{fontenc}
\usepackage{graphicx}
\usepackage[rightcaption]{sidecap}
\usepackage{epsfig, float}
\usepackage{pgf,tikz,pgfplots}
\usetikzlibrary{arrows}
\usepackage{enumitem}
\usepackage{dcolumn}
\pgfplotsset{compat=1.10}
\usepackage{amsmath,amssymb}
\usepackage{amsfonts}
\usepackage{url}
\usepackage{bbm}
\usepackage{mathrsfs} 
\usepackage[english]{babel}
\usepackage[unicode=true, pdfusetitle, bookmarks=true,
  bookmarksnumbered=false, bookmarksopen=false, breaklinks=true, 
  pdfborder={0 0 0}, backref=false, colorlinks=true, linkcolor=blue,
  citecolor=blue, urlcolor=blue]{hyperref}
\usepackage{slashed}
\usepackage{dsfont}
\usepackage{mathabx}
\usepackage{cite}
\usepackage[normalem]{ulem}

\newcommand{\R}{\mathbb{R}}
\newcommand{\CC}{\mathbb{C}}
\newcommand{\N}{\mathbb{N}}

\newcommand{\M}{\mathbb{M}}

\newcommand{\vx}{\mathbf{x}}
\newcommand{\vy}{\mathbf{y}}
\newcommand{\vz}{\mathbf{z}}
\newcommand{\vb}{\mathbf{b}}
\newcommand{\be}{\begin{equation}}
\newcommand{\ee}{\end{equation}}
\newcommand{\ret}{{\rm ret}}
\newcommand{\sym}{{\rm sym}}

\newcommand{\free}{{\rm free}}

\newcommand{\vertiii}[1]{{\left\vert\kern-0.25ex\left\vert\kern-0.25ex\left\vert #1 
    \right\vert\kern-0.25ex\right\vert\kern-0.25ex\right\vert}}
\newcommand{\Banach}{\mathscr{B}}

\DeclareMathOperator*{\esssup}{ess \, sup}
\DeclareMathOperator{\past}{past}
\DeclareMathOperator{\future}{future}
\DeclareMathOperator{\sgn}{sgn}

\DeclareMathOperator{\erfi}{erfi}

\newtheorem{theorem}{Theorem}[section]

\newenvironment{remark}[1][Remark:]{\begin{trivlist}
\item[\hskip \labelsep {\bfseries #1}]}{\end{trivlist}}
\newenvironment{remarks}[1][Remarks:]{\begin{trivlist}
\item[\hskip \labelsep {\bfseries #1}]}{\end{trivlist}}

\title{Singular light cone interactions of scalar particles in 1+3 dimensions}

\author{
Matthias Lienert\thanks{Fachbereich Mathematik,
     Eberhard-Karls-Universit\"at T\"ubingen,
     Auf der Morgenstelle 10, 72076 T\"ubingen, Germany.
     E-mail: matthias.lienert@uni-tuebingen.de}  ~and
Markus N\"oth\thanks{Mathematisches Institut, Ludwig-Maximilians-Universit\"at,
	Theresienstr. 39, 80333 M\"unchen, Germany. E-mail: noeth@math.lmu.de}
}

\date{March 19, 2020}

\begin{document}

\maketitle

\begin{abstract}
\noindent Here we consider an integral equation describing a fixed number of scalar particles which interact not through boson exchange but directly along light cones, similarly as in bound state equations such as the Bethe-Salpeter equation. The equation involves a multi-time wave function $\psi(x_1,...,x_N)$ with $x_i=(t_i,\vx_i) \in \R^4$ as a crucial concept. Assuming a cutoff in time, we prove that it has a unique solution for all data at the initial time. The cutoff is justified by considering the integral equation for a particular curved spacetime with a Big Bang singularity where an initial time occurs naturally without violating any spacetime symmetries. The main feature of our work is that we treat the highly singular case that interactions occur exactly at zero Minkowski distance, reflected by a delta distribution along the light cone. We also extend the existence and uniqueness result to an arbitrary number $N \geq 2$ of particles. Overall, we provide a rigorous example for a certain type of interacting relativistic quantum dynamics in 1+3 spacetime dimensions.
\\[0.1cm]

\end{abstract}

\section{Introduction}
\label{sec:intro}

\subsection{Motivation}
The goal of this paper is to prove the existence and uniqueness of solutions of the equation
\be
	\psi(x,y) = \psi^\free(x,y) + \int d^4 x' \,d^4 y'~G_1^\ret(x-x') G_2^\ret(y-y') K(x',y')\psi(x',y')
\label{eq:inteq} 
\ee
and its $N$-particle generalization for the singular case of light cone interactions, i.e., for
\be
	K(x,y) ~=~ \frac{\lambda}{4\pi} \delta((x-y)^2).
\ee
Here, $\lambda > 0$ is a coupling constant and $(x-y)^2 = (x^0-y^0)^2 - |\vx-\vy|^2$ stands for the Minkowski distance of the spacetime points $x=(x^0,\vx)$ and $y=(y^0,\vy)$. Moreover, $\psi^\free$ is a solution of the free Klein-Gordon equation in each variable, i.e., $(\Box_k+m_k^2)\psi^\free(x_1,x_2) = 0$, $k=1,2$. We shall later see that $\psi^\free$ plays the role of initial data for Eq.\@ \eqref{eq:inteq}. $G_k^\ret$ stands for the retarded Green's function of the respective Klein-Gordon equation. $\psi$ is a \textit{multi-time wave function}, i.e., for $N=2$ particle, a map
\be
	\psi : \text{spacetime} \times \text{spacetime} \rightarrow \CC,~~~(x,y) \mapsto \psi(x,y).
\label{eq:multitimewf}
\ee
The crucial point about Eq.\@ \eqref{eq:inteq} is that it describes a fixed number of (here $N=2$) interacting particles in a manifestly Lorentz invariant way. Interactions happen directly along the light cones instead of through particle exchange as in quantum field theory. Such a relativistically invariant interacting dynamics for a fixed number of particles in 1+3 spacetime dimensions is difficult to achieve; in fact, for Hamiltonian theories this is generally believed to be impossible, and there have long been no-go theorems in that direction (see e.g. \cite{nointeraction,nointeraction2}). It is therefore not surprising that Eq.\@ \eqref{eq:inteq} has a distinctly non-Hamiltonian character. This is evident from the fact that the interaction term involves values of $\psi$ in the past, not only on a Cauchy surface which defines the present. In fact, the time delay of the interaction is an important resource of the kind of dynamics defined by Eq.\@ \eqref{eq:inteq}, and it is only made possible by the more general notion of wave function, the multi-time wave function $\psi$. 

The concept of multi-time wave functions enjoys a long history, going back to well-known physicists such as Eddington \cite{eddington}; Dirac \cite{dirac_32}; Dirac, Fock, Podolsky \cite{dfp}; Bloch \cite{bloch}; Tomonaga \cite{tomonaga} and Schwinger \cite{schwinger}. While it was intermittently picked up during the years (see e.g. \cite{guenther_1952,marx_1974,schweber,drozvincent_1981,sazdjian_2bd,2bdem}, it has recently received renewed attention and undergone significant developments \cite{nogo_potentials,qftmultitime,multitime_pair_creation,1d_model,nt_model,2bd_current_cons,deckert_nickel_2016,lpt_2017b,generalized_born,ibc_model,phd_nickel,deckert_nickel_2019,lnt_2020}; an overview from 2016 can be found in \cite{dice_paper}. The idea is straightforward: to seek a Lorentz covariant generalization of the usual Schr\"odinger picture wave function (here for $N=2$)
\be
	\varphi : \R \times \R^3 \times \R^3 \rightarrow \CC,~~~~~ (t,\vx,\vy)\mapsto \varphi(t,\vx,\vy).
\ee
In fact, the relation of $\psi$ to $\varphi$ is just given by evaluation of $\psi$ at equal times in a given Lorentz frame:
\be
	\varphi(t,\vx,\vy)~=~\psi(t,\vx,t,\vy).
\ee
For the present purposes, the point of interest is that the concept of a multi-time wave function allows us to see the integral equation \eqref{eq:inteq} as the natural generalization of the Schr\"odinger equation $(i \partial_t - H_1^\free - H_2^\free - V(t,\vx,\vy))\varphi(t,\vx,\vy) = 0$ when formulated as an integral equation (see \cite{direct_interaction_quantum} for a more detailed discussion). The latter can namely be written as
\begin{align}
	\varphi(t,\vx,\vy) = \varphi^\free(t,\vx_1,\vx_2) + \int_0^\infty \!\! dt' \int d^3 \vx' \, d^3 \vy' &\, G_1^\ret(t-t',\vx-\vx') G_2^\ret(t-t',\vy-\vy') \nonumber\\
&\times V(t',\vx',\vy')\varphi(t',\vx',\vy'),
	\label{eq:schroedint}
\end{align}
where $\varphi^\free$ is a solution of $(i \partial_t - H_1^\free - H_2^\free)\varphi(t,\vx,\vy) = 0$ and $G_k^\ret$ stands for the retarded Green's function of the operator $(i\partial_t - H_k^\free)$.

Now, our previous integral equation \eqref{eq:inteq} reduces to a very similar equation when we neglect the time delay $|\vx-\vy|$ (we here work with units where $\hbar = 1 = c$) by replacing
\be
	\delta((x'-y')^2) = \frac{1}{2|\vx'-\vy'|}\left[ \delta({x'}^0-{y'}^0-|\vx'-\vy'|) + \delta({x'}^0-{y'}^0+|\vx'-\vy'|) \right]
\label{eq:lcdeltasplit}
\ee
with $\delta({x'}^0-{y'}^0)/|\vx'-\vy'|$. After performing the time integration over ${y'}^0$, renaming the remaining time variable ${x'}^0$ as $t'$ and considering on the left hand side of \eqref{eq:inteq} at equal times $x^0= t = y^0$ in the given Lorentz frame, we arrive at \eqref{eq:schroedint} with Klein-Gordon Green's functions and $V(t,\vx,\vy) \propto 1/|\vx-\vy|$, the Coulomb potential.

This train of thought suggests that Eq.\@ \eqref{eq:inteq} constitutes a natural generalization of the Schr\"odinger equation \eqref{eq:schroedint} for relativistic quantum phenomena (here for two scalar particles with electromagnetic interactions) -- at least for processes where particle creation and annihilation are not relevant, such as relativistic bound states. Interestingly, this is also the domain where the well-known Bethe-Salpeter equation \cite{bs_equation,nakanishi} of quantum field theory is usually applied. The equation also involves a multi-time wave function \eqref{eq:multitimewf} and has a similar form as \eqref{eq:inteq}. The main differences, however, are that for the Bethe-Salpeter equation (a) the interaction kernel $K$ is not given by a clear-cut expression such as $\delta((x-y)^2)$ but by a (potentially divergent) infinite series of Feynman diagrams, and (b) that it involves Feynman Green's functions instead of retarded Green's functions. Contrary to retarded Green's functions, Feynman Green's functions have support not only along and inside backward light cones. Nevertheless, the similarity of \eqref{eq:inteq} with the Bethe-Salpeter equation constitutes further motivation for its study.

In this context, it is interesting to note that previous works about the existence and solutions of simplified models for the Bethe-Salpeter equation (as in the so-called Wick-Cutkosky model, see \cite{wick_54,cutkosky54,green57,consenza65,tiktopoulos65,obrien75}) have (to the best of our knowledge) not answered the question of the existence and uniqueness of solutions. Rather, they omit the free solution $\psi^\free$ from the equation (which, as we will see, leads only to the trivial solution in our case), perform a Fourier transform in all eight variables, assume a plane wave in the center-of-momentum coordinate, use a Wick rotation and then study the eigenvalue problem of a resulting equation of the qualitative form $\widetilde{\psi} = \lambda \widehat{K}(E) \widetilde{\psi}$ in $\lambda$ (instead of the energy $E$ in the center-of-momentum frame). A transformation back to the original problem is not attempted (and may not always be possible). While these results are nevertheless interesting as they reveal features of possible stationary states of the actual problem (i.e., for the physical value of $\lambda$), they are far-removed from the physical problem of time evolution of the quantum-mechanical wave function which we attempt to address here.

\subsection{Previous works}

The physical ideas underlying the multi-time integral equation \eqref{eq:inteq} were first introduced in \cite{direct_interaction_quantum}. That paper includes a more detailed derivation of \eqref{eq:inteq} as a relativistic generalization of the integral version of the Schr\"odinger equation, the treatment of the non-retarded limit as well as a comparison with differential multi-time equations. Furthermore, it discusses the parallels of \eqref{eq:inteq} with classical action-at-a-distance electrodynamics (where interactions also occur directly along the light cone). In addition, different $N$-particle generalizations of \eqref{eq:inteq} are compared and analyzed.

The first rigorous results about the existence and uniqueness of solutions of multi-time integral equations of the form \eqref{eq:inteq} were obtained in \cite{mtve}. This article also focuses on the (simpler) case of scalar particles; in addition, it makes two important assumptions: (i) Only bounded or weakly singular interactions kernels $K$ instead of $\delta((x-y)^2)$ are considered which makes the problem much easier to treat. (ii) A cutoff in time is assumed, meaning that the domain of integration is only $(\frac{1}{2}\M)\times (\frac{1}{2}\M)$ where $\tfrac{1}{2}\M = [0,\infty)\times \R^3$ denotes a Minkowski half-space. This assumption together with the fact that the retarded Green's functions are only supported on and inside of the backward light cone has the important effect of rendering the domain of integration in \eqref{eq:inteq} finite. In addition, one obtains a Volterra-structure in the time variables, meaning that the time integrations in ${x'}^0$ and ${y'}^0$ only run from 0 to $x^0$ and from 0 to $y^0$, respectively. This, in turn, made it possible to utilize an efficient iteration scheme for the proof of existence and uniqueness of solutions. The result was that for every free solution $\psi^\free$ in a suitable Banach space, the integral equation posseses a unique solution $\psi$ in that space which, furthermore, agrees with $\psi^\free$ at the initial time, i.e., for $x^0 = 0 = y^0$.

The assumption of a cutoff in time was made in \cite{mtve} with reference to a potential Big Bang singularity without, however, considering \eqref{eq:inteq} on curved spacetimes. To carry out this task for a class of spacetimes where the Green's functions of the (conformal) wave equation are explicitly known was the topic of \cite{int_eq_curved}. There, it was shown that Eq.\@ \eqref{eq:inteq} has a straightforward generalization to curved spacetimes. A number of explicit examples (flat, open and closed Friedman-Lema\^itre-Robertson-Walker (FLRW) spacetimes) was formulated, and it was shown that for most of these cases, conformal transformations could be used to reduce the proof of existence and uniqueness of solutions to the one on $(\frac{1}{2}\M)^2$. Thereby, the point was made that a cutoff in time can arise naturally in a cosmological context, without violating any spacetime symmetries.

The most recent work about multi-time integral equation is \cite{int_eq_dirac}. It is concerned with extending the previous results to the case of Dirac particles. This has been achieved for a class of sufficiently regular interaction kernels $K(x,y)$. The main difficulty in the Dirac case compared to the Klein-Gordon case is that the Green's functions involve distributional derivatives which complicates the analysis. In particular, it becomes necessary to achieve a delicate balance of the regularity of solutions with the form of the integral equation. Apart from this, the work \cite{int_eq_dirac} also led to some technical developments where the method of proof was refined by directly using a contraction argument on a weighted Sobolev space instead of the Volterra iteration scheme of \cite{mtve,int_eq_curved}.

\subsection{Overview of the paper}

The goal of the present paper is to extend the previous results for scalar particles to the physically most interesting case $K(x,y) \propto \delta((x-y)^2)$. This is, at the same time, a highly singular and therefore challenging case. It becomes necessary to define the particular combination of the three distributions $G^\ret_1(x-x')$, $G_2^\ret(y-y')$ and $\delta((x-y)^2)$ which occurs in \eqref{eq:inteq} and then prove the existence and uniqueness of the resulting singular integral equation. This equation significantly differs from that considered in \cite{mtve} where through admitting only less singular interaction kernels $K$ only two singular distributions acting on different variables needed to be considered.

The paper is structured as follows. In Sec.\@ \ref{sec:formulation} it is shown how to define the integral equation in a rigorous way (by using the delta distributions to eliminate certain integration variables). To this end, we again consider the equation on the Minkowski half-space (assuming a cutoff in time). Section \ref{sec:results} contains our main results: Thm.\@ \ref{thm:bounds} contains explicit bounds for the integral operator in terms of a general weight function of a weighted $L^\infty$ space. Thm.\@ \ref{thm:exponentialg} shows that in the case of massless particles already an exponential weight function leads to the existence and uniqueness of solutions of the integral equation. Our main result is Thm.\@ \ref{thm:existence}, an existence and uniqueness theorem for the full (massive) case. In that case, a different weight function growing like the exponential of a polynomial is used. 

Section \ref{sec:Npart} deals with generalizing this existence and uniqueness theorem to $N$ scalar particles; the corresponding theorem, Thm.\@ \ref{thm:Npart}, is a direct consequence of Thm.\@ \ref{thm:existence}. To the best of our knowledge, this is the first rigorous result about a multi-time integral equation for $N$-particles. 

In Section \ref{sec:curvedspacetime} we show by considering a specific example (an open FLRW spacetime) that the cutoff in time can be achieved naturally for a cosmological spacetime with a Big Bang singularity, without breaking any spacetime symmetries. That is, we show the equivalent result of \cite{int_eq_curved} for singular light cone interactions. The respective existence and uniqueness theorem is Thm.\@ \ref{thm:existencecurved}.

Section \ref{sec:proofs} contains the proofs. In Sec.\@ \ref{sec:conclusions}, we conclude.

\section{Precise formulation of the problem}
\label{sec:formulation}

In the following, we show how to precisely define the integral equation \eqref{eq:inteq} for the case of two scalar particles with masses $m_1$ and $m_2$ on the Minkowski half space $\frac{1}{2}\M = [0,\infty) \times \R^3$. Strictly speaking, to introduce a cutoff in time in this way breaks the Poincar\'e invariance of \eqref{eq:inteq}; however, we will give an argument for its use in Sec.\@ \ref{sec:curvedspacetime}.

It is necessary to take special care of the definition of the integral equation as it contains certain combinations (convolutions and products) of distributions (the Green's functions). Our strategy is to consider the integral operator acting on test functions first where its action can be defined straightforwardly. Later it will be shown that it is bounded on test functions with respect to a suitably chosen weighted norm. This will make it possible to linearly extend the integral operator to the completion of test functions with respect to that norm.

The retarded Green's function of the Klein-Gordon equation is given by:
\be
	G^\ret(x) ~=~ \frac{1}{4\pi|\vx|}\delta(x^0-|\vx|) - \frac{m}{4\pi \sqrt{x^2}} H(x^0-|\vx|) \frac{J_1(m\sqrt{x^2})}{\sqrt{x^2}} 
	\label{eq:gretkg}
\ee
where $H$ denotes the Heaviside function and $J_1$ stands for a Bessel function of the first kind.
Then, with $K(x,y) = \frac{\lambda}{4\pi} \delta((x-y)^2)$, our integral equation \eqref{eq:inteq} on $(\tfrac{1}{2}\M)^2$ becomes:
\be
	\psi ~=~ \psi^\free + A \psi
\label{eq:abstractinteq}
\ee
where $A = A_0 + A_1 + A_2 + A_{12}$ and
\begin{align}
	(A_0 \psi)(x,y) ~&=~ \frac{\lambda}{(4\pi)^3} \int_0^{x^0} d {x'}^0 \int_{\mathbb{R}^3} d^3 \vx' \int_0^{y^0} d{y'}^0 \int_{\mathbb{R}^3} \frac{\delta(x^0-{x'}^0-|\vx-\vx'|)}{|\vx-\vx'|}\nonumber\\ &~~~~\times \frac{\delta(y^0-{y'}^0-|\vy-\vy'|)}{|\vy-\vy'|}\delta((x'-y')^2) \psi(x',y'),\label{eq:a0informal}\\
	(A_1 \psi)(x,y) ~&=~  -\frac{\lambda \, m_1}{(4 \pi)^3} \int_0^\infty d{x'}^0 \int d^3 \vx' \int_0^\infty d{y'}^0 \int d^3 \vy'~H(x^0-{x'}^0-|\vx-\vx'|)  \nonumber\\
& ~~~~\times \frac{J_1(m_1\sqrt{(x-x')^2})}{\sqrt{(x-x')^2}} \frac{\delta(y^0-{y'}^0-|\vy-\vy'|)}{|\vy-\vy'|} \delta((x'-y')^2) \psi(x',y')\label{eq:a1informal}\\
	(A_2 \psi)(x,y) ~&=~ -\frac{\lambda \, m_2}{(4 \pi)^3} \int_0^\infty d{x'}^0 \int d^3 \vx' \int_0^\infty d{y'}^0 \int d^3 \vy'~\frac{\delta(x^0-{x'}^0-|\vx-\vx'|)}{|\vx-\vx'|} \nonumber\\
& ~~~~\times H(y^0-{y'}^0-|\vy-\vy'|) \frac{J_1(m_2\sqrt{(y-y')^2})}{\sqrt{(y-y')^2}} \delta((x'-y')^2) \psi(x',y') \label{eq:a2informal}\\
	(A_{12} \psi)(x,y) ~&=~ \frac{\lambda \, m_1 m_2}{(4\pi)^3}  \int_0^\infty d{x'}^0 \int d^3 \vx' \int_0^\infty d{y'}^0 \int d^3 \vy'~ H(x^0-{x'}^0-|\vx-\vx'|)  \nonumber\\
&~~~~\times \frac{J_1(m_1\sqrt{(x-x')^2})}{\sqrt{(x-x')^2}} H(y^0-{y'}^0-|\vy-\vy'|) \frac{J_1(m_2\sqrt{(y-y')^2})}{\sqrt{(y-y')^2}}\nonumber\\
&~~~~\times \delta((x'-y')^2) \psi(x',y')\label{eq:a12informal}.
\end{align}
We now formally manipulate these informal expressions in such a way that the end results can be given a precise meaning on test functions. Let $\mathcal{S} = \mathcal{S}((\tfrac{1}{2}\M)^2)$ denote the space of Schwartz functions on $(\tfrac{1}{2}\M)^2$, and let $\psi \in \mathcal{S}$. 

\paragraph{Definition of $A_0$.}
We consider the massless term $A_0$ first which is also the most singular term. Using the $\delta$-functions to eliminate the integration over ${x'}^0$ and ${y'}^0$ results in:
\begin{align}
	(A_0 \psi)(x,y) = \frac{\lambda}{(4\pi)^3} \int_{B_{x^0}(\vx)} \hspace{-0.5cm}d^3\vx' \int_{B_{y^0}(\vy)} \hspace{-0,5cm}d^3 \vy' \,  \frac{\delta((x^0-y^0-|\vx'|+|\vy'|)^2-|\vx-\vy +\vx'-\vy'|^2)}{|\vx'||\vy'|} \nonumber \\
    \psi(x+x',y+y')|_{{x'}^0 = -|\vx'|, \, {y'}^0 = -|\vy'|},
	\label{eq:a0informal2}
\end{align}
Note that the domain of integration has been reduced to a compact region whose size depends on $x^0$ and $y^0$. There is still one more $\delta$-distribution left. We choose to use it to eliminate $|\vx'| =: r$. It is convenient to introduce the vector
\be
	b = x-y-(-|\vy'|, \vy').
\label{eq:b}
\ee
Then the argument of the delta function can be written as:
\be
	(b^0-|\vx'|)^2 - |\vb + \vx'|^2.
\ee
This expression has a root in $r$ for
\be   
	r =  r^* := \frac{1}{2} \frac{b^2}{b^0 + |\vb| \cos \vartheta}
\label{eq:r}
\ee
where $\vartheta$ is the angle between $\vb$ and $\vx'$. Of course, $r^*$ inherits the restrictions of the range of $r$, thus is only a valid root for
\be
	0 < r^* < x^0.
\ee
The requirement $0 < r^*$ can be satisfied in two cases, either \(b^2>0\) and \(b^0>0\), or \(b^2<0\) and \(\cos\vartheta< - \frac{b^0}{|\vb|}\). Using these restrictions, the condition $r^*< x^0$ can be converted into a restriction of the domain of integration in \(\vartheta\):
\begin{align}
   & \frac{1}{2}\frac{b^2}{b^0+|\vb|\cos\vartheta}~ <~x^0\nonumber\\
   \iff~~~ &\sgn(b^2) b^2 ~< ~ 2x^0 \sgn(b^2) ( b^0+|\vb|\cos \vartheta)\nonumber\\
    \iff~~~ &\frac{|b^2|}{2x^0 |\vb|} -\frac{\sgn(b^2)b^0}{|\vb|} ~<~ \sgn(b^2) \cos \vartheta\nonumber\\
    \iff ~~~&\left\{\begin{matrix}\cos\vartheta > \frac{b^2}{2x^0|\vb|}- \frac{b^0}{|\vb|}, \quad \text{for} ~ b^2 >0 \\ \cos\vartheta < \frac{b^2}{2x^0|\vb|} - \frac{b^0}{|\vb|}, \quad \,  \text{for} ~ b^2<0. \end{matrix} \right.
\end{align}
In case of $b^2<0$, the new restriction on \(\cos\vartheta\) is stricter than \(\cos\vartheta<-\frac{b^0}{|\vb|}\); we thus use it to replace the latter. We evaluate the $\delta$-function using spherical coordinates in $\vy'$ and the usual rule
\begin{equation}
    \delta(f(z))=\sum_{z^* : f(z^*)=0} \frac{\delta(z-z^*)}{|f'(z^*)|},
\end{equation}
where \(f(r)=(b^0-r)^2-(\vb+x')^2= -(r-r^*)2(b^0+|\vb|\cos\vartheta) \). 
The result is an expression for \(A_0\psi\) which does not contain distributions anymore:
\begin{align}\nonumber
    &(A_0\psi)(x,y)=\frac{\lambda}{(4\pi)^3}\int_{B_{y^0}(\vy)}d^3\vy'  \int_0^{2\pi}d\varphi \int_{-1}^{1} d\!\cos\vartheta ~\frac{|b^2|}{4(b^0+|\vb|\cos\vartheta)^2 |\vy'|} \psi(x+x',y+y')\\
    &\left(1_{b^2>0}1_{b^0>0} 1_{\cos\vartheta > \frac{b^2}{2x^0|\vb|} - \frac{b^0}{|\vb|}}+1_{b^2<0}1_{\cos\vartheta<\frac{b^2}{2x^0|\vb|} - \frac{b^0}{|\vb|}}\right),
\label{eq:defa0}
\end{align}
still subject to \(x'^0=-r^*=-|\vx'| , {y'}^0=-|\vy'|\). The different cases for $b$ have been implemented through the various indicator functions. Eq.\@ \eqref{eq:defa0} will serve as our \textit{definition} of $A_0$ on test functions $\psi \in \mathcal{S}$.

\paragraph{Definition of $A_1$.}
Next, we turn to the definition of $A_1$, starting from the informal expression \eqref{eq:a1informal}. We first split up the $\delta$-function of the interaction kernel according to \eqref{eq:lcdeltasplit}.
Then we use $\delta(y^0-{y'}^0-|\vy-\vy'|)$ to eliminate ${y'}^0~(=y^0-|\vy-\vy'|)$. Note that the order of these two steps does not matter. This yields:
\begin{align}
	&(A_1 \psi)(x,y) =  -\frac{\lambda \, m_1}{2(4 \pi)^3} \int_0^\infty d{x'}^0 \int d^3 \vx' \int d^3 \vy'~H(x^0-{x'}^0-|\vx-\vx'|)  \nonumber\\
& ~~~\times \frac{J_1(m_1\sqrt{(x-x')^2})}{\sqrt{(x-x')^2}} \frac{H(y^0-|\vy-\vy'|)}{|\vy-\vy'|} \frac{1}{|\vx'-\vy'|} \left[\delta({x'}^0 - y^0 + |\vy-\vy'| - |\vx'-\vy'|) \right.\nonumber\\
&~~~\left.+ \, \delta({x'}^0 - y^0 +|\vy-\vy'| + |\vx'-\vy'|) \right] \psi(x',y^0-|\vy-\vy'|,\vy').\label{eq:a1informal2}
\end{align}
Finally, we use the remaining $\delta$-functions to eliminate ${x'}^0$. We obtain:
\begin{align}
	&(A_1 \psi)(x,y) =  -\frac{\lambda \, m_1}{2(4 \pi)^3} \int d^3 \vx' \int d^3 \vy'~\frac{H(y^0-|\vy-\vy'|)}{|\vy-\vy'|} \frac{1}{|\vx'-\vy'|} \nonumber\\
&\left[ H({x'}^0)H(x^0-{x'}^0-|\vx-\vx'|)   \left.\frac{J_1(m_1\sqrt{(x-x')^2})}{\sqrt{(x-x')^2}} \psi(x',y')\right|_{\substack{{y'}^0 = y^0-|\vy-\vy'|,\\{x'}^0 = y^0 - |\vy-\vy'| + |\vx'-\vy'| }}\right.\nonumber\\
&\left.+ \, H({x'}^0)H(x^0-{x'}^0-|\vx-\vx'|)   \left.\frac{J_1(m_1\sqrt{(x-x')^2})}{\sqrt{(x-x')^2}} \psi(x',y')\right|_{\substack{{y'}^0 = y^0-|\vy-\vy'|,\\{x'}^0 = y^0 - |\vy-\vy'| - |\vx'-\vy'| }} \right] .\label{eq:defa1}
\end{align}
This expression is free of distributions, so it will serve as our definition of $A_1$ on test functions $\psi \in \mathcal{S}$. Note that the domain of integration is effectively finite due to the Heaviside functions.

\paragraph{Definition of $A_2$.} Starting from \eqref{eq:a2informal}, the analogous steps as for $A_1$ yield:
\begin{align}
	&(A_2 \psi)(x,y) =  -\frac{\lambda \, m_2}{2(4 \pi)^3} \int d^3 \vx' \int d^3 \vy'~\frac{H(x^0-|\vx-\vx'|)}{|\vx-\vx'|} \frac{1}{|\vx'-\vy'|} \nonumber\\
&\left[ H({y'}^0)H(y^0-{y'}^0-|\vy-\vy'|)   \left.\frac{J_1(m_2\sqrt{(y-y')^2})}{\sqrt{(y-y')^2}} \psi(x',y')\right|_{\substack{{x'}^0 = x^0-|\vx-\vx'|,\\{y'}^0 = x^0 - |\vx-\vx'| + |\vx'-\vy'| }}\right.\nonumber\\
&\left.+ \, H({y'}^0)H(y^0-{y'}^0-|\vy-\vy'|)   \left.\frac{J_1(m_2\sqrt{(y-y')^2})}{\sqrt{(y-y')^2}} \psi(x',y')\right|_{\substack{{x'}^0 = x^0-|\vx-\vx'|,\\{y'}^0 = x^0 - |\vx-\vx'| - |\vx'-\vy'| }} \right] .\label{eq:defa2}
\end{align}
This serves as our definition of $A_2$ on test functions $\psi \in \mathcal{S}$.

\paragraph{Definition of $A_{12}$.} Here, we start with \eqref{eq:a12informal}. We change variables $(\vx',\vy') \rightarrow (\vx',\vz = \vx'-\vy')$ (Jacobi determinant $=1$), with the goal of using the remaining $\delta$-function to eliminate $|\vz| = |\vx'-\vy'|$ in mind. We find:
\begin{align}
(A_{12} \psi)(x,y) &= \frac{\lambda \, m_1 m_2}{(4\pi)^3}  \int_0^\infty d{x'}^0 \int d^3 \vx' \int_0^\infty d{y'}^0 \int d^3 \vz~ H(x^0-{x'}^0-|\vx-\vx'|)  \nonumber\\
&~~~\times \frac{J_1(m_1\sqrt{(x-x')^2})}{\sqrt{(x-x')^2}} H(y^0-{y'}^0-|\vy-\vx'+\vz|) \frac{J_1(m_2\sqrt{(y-y')^2})}{\sqrt{(y-y')^2}}\nonumber\\
&~~~\times \delta(({x'}^0-{y'}^0)^2 -|\vz|^2) \psi(x',y')\Big|_{\vy' = \vx'-\vz}\label{eq:a12informal2}.
\end{align}
Now we use spherical coordinates for $\vz$ and eliminate $|\vz|$ through the $\delta$-function, using
\be
	 \delta(({x'}^0-{y'}^0)^2 -|\vz|^2) =  \frac{1}{2 |\vz|} \delta(|{x^0}'-{y^0}'|-|\vz|).
\ee
This yields:
\begin{align}
&(A_{12} \psi)(x,y) = \frac{\lambda \, m_1 m_2}{2(4\pi)^3}  \int_0^\infty d{x'}^0 \int d^3 \vx' \int_0^\infty d{y'}^0 \int_0^{2\pi} d\varphi \int_{0}^{\pi} d \vartheta \, \cos(\vartheta) |{x'}^0-{y'}^0| \,  \nonumber\\
&~~~\times H(x^0-{x'}^0-|\vx-\vx'|) \frac{J_1(m_1\sqrt{(x-x')^2})}{\sqrt{(x-x')^2}}\nonumber\\
&~~~\times H(y^0-{y'}^0-|\vy-\vx'+\vz|) \frac{J_1(m_2\sqrt{(y-y')^2})}{\sqrt{(y-y')^2}} \psi(x',y')\Big|_{\vy' = \vx'-\vz, \, |\vz| = |{x^0}'-{y^0}'|}
\label{eq:defa12}.
\end{align}
The resulting expression does not contain distributions anymore and will serve as our definition of $A_{12}$ on test functions $\psi \in \mathcal{S}$. Note that the domain of integration is again effectively finite.

\paragraph{Lifting of the integral operator from test functions to a suitable Banach space.}
In order to prove the existence and uniqueness of solutions of the integral equation $\psi = \psi^\free + A\psi$, we need to define the operator $A$ not only on test functions but on a suitable Banach space which includes (at least) sufficiently many solutions $\psi^\free$ of the free multi-time Klein-Gordon equations, $(\Box_k + m_k^2)\psi^\free(x_1,x_2) = 0,~k=1,2$. 
We shall define this Banach space as the completion of $\mathcal{S} =\mathcal{S}((\tfrac{1}{2}\M)^2)$ with respect to a suitable norm. A good choice which works well for the upcoming existence and uniqueness proofs is the class of weighted $L^\infty$-norms
\be
	\| \psi \|_g := \esssup_{x,y \in \tfrac{1}{2}\M} \frac{|\psi(x,y)|}{g(x^0)g(y^0)},
\ee
where $g : \R^+_0 \rightarrow \R^+$ is assumed to be a monotonically increasing function such that $1/g$ is bounded. Then our Banach space is given by the completion
\be
	\Banach_g = \overline{\mathcal{S}}^{\| \cdot \|_g}.
\ee
Our next goal is to find a weight function $g$ such that the operator $A$ is not only bounded but even defines a contraction on $\Banach_g$. By linear extension, it is sufficient to estimate $\| A \psi \|_g$ on test functions $\psi \in \mathcal{S}$.

\begin{remarks}
	\begin{enumerate}
		\item We have attempted to use an $L^\infty_t L^2_\vx$-based norm ($L^\infty$ in the times and $L^2$ in the space variables). However, we did not succeed with obtaining suitable estimates for that case. This might not be a problem in principle, but its treatment would require further technical innovation.
More precisely, one would need to understand integral operators such as \eqref{eq:defa0} whose kernel is in $L^1$ but not in $L^2$.
		\item Nevertheless, our definition of $\Banach_g$ contains a large class of free solutions of the Klein-Gordon equation. As the Klein-Gordon equation preserves boundedness, all bounded initial data for $\psi^\free$ lead to a free solution $\psi^\free \in \Banach_g$ which can be used as an input to our integral equation.
	\end{enumerate}
\end{remarks}

\section{Results}
\label{sec:results}

This section is structured as follows. Sec.\@ \ref{sec:2part} (which is about the two-particle case) contains the main results: the estimates of the integral operators as well as the theorems about existence and uniqueness of solutions. Sec.\@ \ref{sec:Npart} we extend these results to the $N$-particle case and in Sec.\@ \ref{sec:curvedspacetime} we show that a curved spacetime with a Big Bang singularity can provide a natural reason for a cutoff in time.

\subsection{The two-particle case} \label{sec:2part}

For $t\geq0$, we define the functions:
\begin{align}
	g_0(t) &= g(t),\nonumber\\
\text{and for }n\in\N:~~~	g_n(t) &= \int_0^t dt' \, g_{n-1}(t').	
\end{align}
Note that due to the properties of $g$, the functions $g_n$ are monotonically increasing for all $n\in \N$; furthermore, by definition, they satisfy $g_n(0)=0$.

Our first theorem gives explicit bounds for the operators $A_0, A_1, A_2, A_{12}$ in terms of the functions $g_n$. The proof can be found in Sec.\@ \ref{sec:proofbounds}.

\begin{theorem}[Bounds of the integral operators on $\mathcal{S}$.]
	\label{thm:bounds}
	For all $\psi \in \mathcal{S}((\tfrac{1}{2}\M)^2)$, the integral operators $A_0, A_1, A_2, A_{12}$ satisfy the following bounds:
	\begin{align}
		\sup_{\psi \in \mathcal{S}((\frac{1}{2}\M)^2)} \frac{\| A_0 \psi \|_g}{\| \psi \|_g} ~&\leq~ \frac{\lambda}{8\pi} \left(\sup_{t\geq 0} \frac{g_1(t)}{g(t)}\right)^2\label{eq:estimatea0},\\
	\sup_{\psi \in \mathcal{S}((\frac{1}{2}\M)^2)} \frac{\| A_1 \psi \|_g}{\| \psi \|_g} ~&\leq~ \frac{\lambda \, m_1^2}{16\pi} \left[ 3\left(\sup_{t\geq 0}\frac{t g_1(t)}{g(t)}\right)\left(\sup_{t\geq 0}\frac{g_2(t)}{g(t)}\right) + 3\left( \sup_{t\geq 0} \frac{g_1(t)}{g(t)} \right)\left( \sup_{t\geq 0} \frac{tg_2(t)}{g(t)} \right) \right.\nonumber\\
	&~~~~~\left. +\, 2 \left( \sup_{t\geq 0} \frac{g_1(t)}{g(t)} \right)\left( \sup_{t\geq 0} \frac{g_3(t)}{g(t)} \right) \right], \label{eq:estimatea1}\\
		\sup_{\psi \in \mathcal{S}((\frac{1}{2}\M)^2)} \frac{\| A_2 \psi \|_g}{\| \psi \|_g} ~&\leq~ \frac{\lambda \, m_2^2}{16\pi} \left[ 3\left(\sup_{t\geq 0}\frac{t g_1(t)}{g(t)}\right)\left(\sup_{t\geq 0}\frac{g_2(t)}{g(t)}\right) + 3\left( \sup_{t\geq 0} \frac{g_1(t)}{g(t)} \right)\left( \sup_{t\geq 0} \frac{tg_2(t)}{g(t)} \right) \right.\nonumber\\
	&~~~~~\left. + \, 2 \left( \sup_{t\geq 0} \frac{g_1(t)}{g(t)} \right)\left( \sup_{t\geq 0} \frac{g_3(t)}{g(t)} \right) \right], \label{eq:estimatea2}\\
	\sup_{\psi \in \mathcal{S}((\frac{1}{2}\M)^2)} \frac{\| A_{12} \psi \|_g}{\| \psi \|_g} ~&\leq~ \frac{\lambda \, m_1^2 m_2^2}{96\pi} \left[ \left( \sup_{t\geq 0} \frac{t^2 g_2(t)}{g(t)}\right)\left( \sup_{t\geq 0} \frac{t g_1(t)}{g(t)}\right) \right.\nonumber\\
&~~~~~\left.+ \, \frac{1}{2}  \left( \sup_{t\geq 0} \frac{t^2 g_3(t)}{g(t)}\right)\left( \sup_{t\geq 0} \frac{g_1(t)}{g(t)}\right) \right].\label{eq:estimatea12}
	\end{align}
\end{theorem}
 In case these expressions are finite, $A_0, A_1, A_2, A_{12}$ extend to linear operators on $\Banach_g$ with the same norms. Our next task is to find suitable weight functions $g$ such that this is actually the case. We begin with the massless case where already an exponential weight function leads to an estimate which remains finite after taking the supremum. The massive case is treated subsequently; it is a little more difficult as all the estimates for the operators $A_0, A_1, A_2, A_{12}$ have to be finite at the same time. This  requires a different choice of weight function (see Thm.\@ \ref{thm:existence}).

\begin{theorem}[Bounds for $A_0$ and $g(t) = e^{\gamma t}$; existence of massless dynamics.]
	\label{thm:exponentialg}
	~\\ For any $\gamma> 0$, let $g(t) = e^{\gamma t}$. Then $A_0$ can be linearly extended to a bounded operator on $\Banach_g$ with norm
	\be
		\| A_0 \| ~\leq~ \frac{\lambda}{8\pi \gamma^2}.
	\label{eq:norma0exponential}
	\ee
	Consequently, for all $\gamma > \sqrt{\frac{\lambda}{8\pi}}$, the integral equation $\psi = \psi^\free + A_0\psi$ has a unique solution $\psi \in \Banach_g$ for every $\psi^\free \in \Banach_g$.
\end{theorem}

Now we come to our main result.

\begin{theorem}[Existence of dynamics in the massive case.]
	\label{thm:existence}~\\
	For any $\alpha > 0$, let
	\be
		g(t) = (1+\alpha t^2)e^{\alpha t^2/2}.
	\label{eq:weightfactor}
	\ee
	 Then $A_0, A_1, A_2$ and $A_{12}$ can be linearly extended to bounded operators on $\Banach_g$ with norms
	\begin{align}
		\| A_0 \| ~&\leq~ \frac{\lambda}{32 \pi} \frac{1}{\alpha}, \label{eq:estimatea0final}\\
		\| A_1 \| ~&\leq~ \frac{5\lambda \, m_1^2}{16\pi} \frac{1}{\alpha^2},\label{eq:estimatea1final}\\
		\| A_2 \| ~&\leq~ \frac{5\lambda \, m_2^2}{16\pi} \frac{1}{\alpha^2}, \label{eq:estimatea2final}\\
		\| A_{12} \| ~&\leq~ \frac{\lambda \, m_1^2 m_2^2}{80\pi} \frac{1}{\alpha^3}. \label{eq:estimatea12final}
	\end{align}
	Consequently, for all $\alpha > 0$ with
\be
	\frac{\lambda}{8\pi \alpha} \left( \frac{1}{4} + \frac{5(m_1^2 + m_2^2)}{2} \frac{1}{\alpha} + \frac{m_1^2 \, m_2^2}{10} \frac{1}{\alpha^2} \right) ~<~ 1,
\label{eq:condexistencemassive}
\ee
	the integral equation $\psi = \psi^\free + A\psi$ has a unique solution $\psi \in \Banach_g$ for every $\psi^\free \in \Banach_g$.
\end{theorem}

The proof can be found in Sec.\@ \ref{sec:proofexistence}.

\begin{remarks}
	\begin{enumerate}
		\item \textit{Comparison of Thms. \ref{thm:exponentialg} and \ref{thm:existence} in the massless case.} On the first glance, the result of Thm.\@ \ref{thm:exponentialg} looks stronger in the sense that for $g(t)=e^{\gamma t}$, the estimate of $\|A_0\|$ goes with $\gamma^{-2}$ while for $g(t)=(1+\alpha t^2 )e^{\alpha t^2/2}$, the estimate of $\|A_0\|$ goes with $\alpha^{-1}$. However, one should note that $\gamma$ is the constant in front of $t$ while $\alpha$ occurs in combination with $t^2$. Thus, if one wants to draw a comparison between these different cases at all, then it should be between $\gamma$ and $\sqrt{\alpha}$. Of course, the main difference between the two theorems is the admitted growth rate of the solutions. In this regard, Thm.\@ \ref{thm:exponentialg} contains the stronger statement.
		\item A \textit{physically realistic value of $\lambda$} is $\frac{1}{137}$, the value of the fine structure constant. In that case, $\alpha$ need not even be particularly large in order for condition \eqref{eq:condexistencemassive} to be satisfied.
		\item \textit{Initial value problem.} By the integral equation \eqref{eq:inteq}, we obtain that the solution $\psi$ satisfies $\psi(0,\vx,0,\vy) = \psi^\free(0,\vx,0,\vy)$. If $\psi^\free$ is a solution of the free multi-time Klein-Gordon equations, then it is itself determined by initial data at $x_1^0,x_2^0=0$. (As the Klein-Gordon equation is of second order in time, these initial data include data for $\partial_{x^0}\psi$, $\partial_{y^0}\psi$ and $\partial_{x^0} \partial_{y^0} \psi$, see \cite[chap. 5]{phd_nickel}.) Thus, we find that $\psi$ is determined by these data at $x_1^0,x_2^0=0$ as well. Note that for later times, $\psi$ and $\psi^\free$ do not, in general, coincide and consequently a similar statement does not hold.
	\item \textit{Finite propagation speed.} The theorem implies that $\psi = \sum_{k=0}^\infty A^k\psi^\free$. As $(A \psi^\free)(x,y)$ involves only values of $\psi^\free$ in $\past(x) \times \past(y)$ where $\past(x)$ denotes the causal past of $x \in \frac{1}{2}\M$ (see Eqs.\@ \eqref{eq:defa0}, \eqref{eq:defa1}, \eqref{eq:defa2}, \eqref{eq:defa12}), so do $A^k \psi^\free$ for all $k \in \N$ and $\psi$. Therefore, we obtain: if the initial data for $\psi^\free$ at $x^0 = 0 = y^0$ are compactly supported in a region $R \subset \left( \{ 0\} \times \R^3\right)^2$, then for all Cauchy surfaces $\Sigma \subset \frac{1}{2}\M$, $\psi|_{\Sigma \times \Sigma}$ is supported in the causally grown set  $\text{Gr}(R,\Sigma) = \left(\bigcup_{(x,y)\in R} \future(x) \times \future(y) \right) \cap (\Sigma \times \Sigma)$ where $\future(x)$ stands for the causal future of $x \in \frac{1}{2}\M$.
	\item \textit{Square integrable solutions.} As a consequence of the previous item, compactly supported and bounded initial data for $\psi^\free$ lead to a compactly supported and bounded solution $\psi$. In particular, this implies that $\psi(x^0,\cdot,y^0)$ lies in $L^2(\R^6)$ for all times $x^0,y^0\geq 0$.
	\end{enumerate}
\end{remarks}

\subsection{The $N$-particle case} \label{sec:Npart}

Here we extend Thm.\@ \ref{thm:existence} from two to $N\geq 3$ scalar particles. While there are different possibilities to generalize the two-particle integral equation \eqref{eq:inteq}, we focus on the one advocated in \cite{direct_interaction_quantum} as the most promising. For
\be
	\psi : \big(\tfrac{1}{2}\M\big)^N \rightarrow \CC,~~~~~(x_1,...,x_N) \mapsto \psi(x_1,...,x_N)
\ee
we consider the integral equation
\begin{align}
	\psi(x_1,...,x_N) ~=~ &\psi^\free(x_1,...,x_N) +\frac{ \lambda}{4\pi} \sum_{i,j =1,...,N; \, i<j} \int_{\tfrac{1}{2}\M} d^4 x_i \int_{\tfrac{1}{2}\M} d^4 x_j~G^\ret_i(x_i-x_i') \nonumber\\
&\times ~G^\ret(x_j-x_j')\delta((x_i'-x_j')^2) \psi(x_1,...,x_i, ...,x_j,..., x_N).
	\label{eq:npartint}
\end{align}
Here, $\psi^\free$ is again a solution of the free Klein-Gordon equations $(\Box_k + m_k^2)\phi(x_k)$ in each spacetime variable and $G^\ret_k$ stands for the retarded Green's function of the operator $(\Box_k + m_k^2)$, $k=1,2,...,N$.

Eq.\@ \eqref{eq:npartint} is written down in an informal way. To define a rigorous version, let $\psi \in \mathcal{S}\big( (\tfrac{1}{2}\M)^N\big)$ be a test function. Moreover, let $A^{(ij)}$ be the integral operator of the two-particle problem acting on the variables $x_i$ and $x_j$ instead of $x = x_1$ and $y=x_2$. We define
\be
	\,^{(N)}\! A ~=~ \sum_{i,j =1,...,N; \, i<j} A^{(ij)}.
\ee
As will be shown below, $\! \,^{(N)}\! A$ can be linearly extended to a bounded operator on the Banach space $\! \,^{(N)}\!\Banach_g$. That space is defined as the completion of $\mathcal{S}\big( (\tfrac{1}{2}\M)^N\big)$ with respect to the norm
\be
	\|\psi \|_g ~=~ \esssup_{x_1,...,x_N \in \frac{1}{2}\M} \frac{|\psi|(x_1,...,x_N)}{g(x_1^0)\cdots g(x_N^0)},
\ee
where the function $g$ is defined as before.

Then we take the equation
\be
	\,^{(N)}\! A ~=~\psi^\free + \! \,^{(N)}\! A \psi.
\label{eq:npartintabstract}
\ee
to be the rigorous version of \eqref{eq:npartint} on $\! \,^{(N)}\!\Banach_g$.

With these preparations, we are ready to formulate the $N$-particle existence and uniqueness theorem.

\begin{theorem}[Existence of dynamics for $N$ particles.]
	\label{thm:Npart}~\\
		For any $\alpha > 0$, let $g(t) = (1+\alpha t^2)e^{\alpha t^2/2}$. Then the operator $\! \,^{(N)}\! A$ can be linearly extended to a bounded operator on $\! \,^{(N)}\!\Banach_g$ with norm
	\be
		\|\! \,^{(N)}\! A\| ~\leq~\frac{\lambda}{8\pi \alpha} \sum_{i,j =1,...,N; \, i<j}  \left( \frac{1}{4} + \frac{5(m_i^2 + m_j^2)}{2} \frac{1}{\alpha} + \frac{m_i^2 \, m_j^2}{10} \frac{1}{\alpha^2} \right).
	\ee
	If $\alpha > 0$ is such that this expression is strictly smaller than one, the integral equation \eqref{eq:npartintabstract} has a unique solution $\psi \in \! \,^{(N)}\!\Banach_g$ for every $\psi^\free \in \! \,^{(N)}\!\Banach_g$.
\end{theorem}

The proof follows straightforwardly from that of Thm.\@ \ref{thm:existence} using
\be
	\|\! \,^{(N)}\! A\| \leq \sum_{i,j =1,...,N; \, i<j} \| A^{(ij)}\|_g.
\ee
For the norms of the operators $A^{(ij)}$, one can use the previous expressions as these operators act only as the identity on variables $x_k$ with $k \notin \{i,j\}$.

\begin{remark}
	To the best of our knowledge, Thm.\@ \ref{thm:Npart} is the first result about the existence and uniqueness of solutions of multi-time integral equations for $N$ particles. While for the present contraction argument the generalization to $N$ particles has been straightforward, this is not the case for other works. For example, the Volterra iterations used in \cite{mtve} become increasingly complicated with increasing particle number $N$. For Dirac particles,  a similar technique as ours was used in \cite{int_eq_dirac}. However, as the Dirac Green's functions contain distributional derivatives, one has to control weak derivatives of the solutions, and the number of such derivatives depends on $N$. That situation also does not allow for such a straightforward generalization to $N$ particles as has been possible here.
\end{remark}

\subsection{On the possible origin of a cutoff in time}
\label{sec:curvedspacetime}

So far, we have assumed a cutoff in time. In the way this has been treated so far, this cutoff breaks the manifest Poincar\'e invariance of our integral equation. In this section, we demonstrate at a particular (simple and tractable) example that such a cutoff can arise naturally if the considered spacetime has a Big Bang singularity. Then the Big Bang defines the initial time. To consider a simple example is necessary as otherwise the Green's functions may not be known in detail, and in that case it would not be possible to explicitly define the integral operator, let alone to carry out an analysis of that operator comparable to the one before.

Our example consists of two massless scalar particles which, in absence of interactions, obey the conformally invariant wave equation on a curved spacetime $\mathcal{M}$ with metric g,
\be
	\left( \Box_g - \xi R \right) \chi = 0,
	\label{eq:conformalwaveeq}
\ee
where $R$ denotes the Ricci scalar and in 1+3 dimensions $\xi = \frac{1}{6}$.

We consider these particles on a flat Friedman-Lema\^itre-Robertson-Walker (FLRW) spacetime which is described by the metric
\be
	ds^2 = a^2(\eta) \left( d\eta^2 - dr^2 - r^2 d \Omega^2 \right),
\ee
where $\eta$ denotes conformal time, $d \Omega$ denotes the surface measure on $\mathbb{S}^2$ and $a(\eta)$ is the so-called \textit{scale function}, a continuous function with $a(0) = 0$ and $a(\eta) > 0$ for $\eta >0$. This form makes it obvious that the spacetime is conformally equivalent to a Minkowski half space $\tfrac{1}{2}\M$, with conformal factor $a(\eta)$.

In this case, it is well-known that the Green's functions of \eqref{eq:conformalwaveeq} on the flat FLRW spacetime $\mathcal{M}$ can be obtained from those of the usual wave equation on $\tfrac{1}{2}\M$ as follows (using coordinates $x=(\eta,\vx)$ and $x'=(\eta',\vx')$ with $\eta,\eta' \in [0,\infty)$ and $\vx,\vx' \in \R^3$; see \cite{int_eq_curved} for a more detailed explanation):
\be
	G_{\mathcal{M}}(x,x') ~=~ \frac{1}{a(\eta)} \frac{1}{a(\eta')} G_{\frac{1}{2}\M}(x,x').
\ee
Inserting the well-known expression for the retarded and symmetric Green's functions on $\tfrac{1}{2}\M$ (see \eqref{eq:gretkg}) yields:
\begin{align}
	G_{\mathcal{M}}^\ret(x,x') ~&=~ \frac{1}{4\pi} \frac{1}{a(\eta) a(\eta')} \frac{\delta(\eta - \eta' -|\vx-\vx'|)}{|\vx-\vx'|}\nonumber\\
G_{\mathcal{M}}^\sym(x,x') ~&=~ \frac{1}{4\pi} \frac{1}{a(\eta) a(\eta')} \delta((\eta-\eta')^2-|\vx-\vx'|^2).
\end{align}

With this information, we are ready to write down the integral equation on $\mathcal{M}$. The generalization of \eqref{eq:inteq} to curved spacetimes is straightforward: $\psi$ becomes a scalar function on $\mathcal{M}\times\mathcal{M}$, one exchanges the Minkowski spacetime volume elements with the invariant 4-volume elements on $\mathcal{M}$, and the Green's functions on $\tfrac{1}{2}\M$ get replaced with those on $\mathcal{M}$ as well. As in the Minkowski case, the interaction kernel is given by the symmetric Green's function. With this, the relevant integral equation becomes:
\be
	\psi(x,y) = \psi^\free(x,y) + \lambda \int_{\mathcal{M}\times \mathcal{M}} dV(x) \, dV(y)~G_1^\ret(x,x')G_2^\ret(y,y') G^\sym(x',y') \psi(x',y'),
	\label{eq:inteqcurved}
\ee
For regular and only weakly singular interaction kernels $K(x,y)$ instead of $G^\sym(x',y')$, the problem of existence and uniqueness of solutions of this equation has been treated in \cite{int_eq_curved} for flat, open and closed FLRW universes; the case of Dirac particles and smooth interaction kernels has been addressed in \cite{int_eq_dirac}. For flat FLRW universes and scalar particles, we here extend \cite{int_eq_curved} to the physically most interesting and mathematically challenging case $K(x,y)=G^\sym(x,y)$.

We now formulate \eqref{eq:inteqcurved} explicitly. The spacetime volume element is given by:
\be
	dV(x) = a^4(\eta) \, d\eta \, d^3 \vx.
\ee
With this information, \eqref{eq:inteqcurved} becomes:
\begin{align}
	\psi(\eta_1,\vx_1,\eta_2,\vx_2) =~ &\psi^\free(\eta_1,\vx_1,\eta_2,\vx_2) + \frac{\lambda}{(4\pi)^3} \frac{1}{a(\eta_1) a(\eta_2)} \int_0^{\eta_1} d \eta_1' \int d^3 \vx_1' \int_0^{\eta_2} d \eta_2' \int d^3  \vx_2'\nonumber\\
&\times ~ a^2(\eta_1') a^2(\eta_2') \frac{\delta(\eta_1-\eta_1' - |\vx_1-\vx_1'|)}{|\vx_1-\vx_1'|}\frac{\delta(\eta_2-\eta_2' - |\vx_2-\vx_2'|)}{|\vx_2-\vx_2'|}\nonumber\\
&\times ~ \delta((\eta_1'-\eta_2')^2-|\vx_1'-\vx_2'|^2) \psi(\eta_1',\vx_1',\eta_2',\vx_2').
\label{eq:inteqcurvedexplicit}
\end{align}
Now let
\be
	\chi(\eta_1,\vx_1,\eta_2) = a(\eta_1) a(\eta_2) \psi(\eta_1,\vx_1,\eta_2).
\ee
and $\chi^\free(\eta_1,\vx_1,\eta_2) = a(\eta_1) a(\eta_2) \psi^\free(\eta_1,\vx_1,\eta_2)$.
Then \eqref{eq:inteqcurvedexplicit} is equivalent to:
\begin{align}
	\chi(\eta_1,\vx_1,\eta_2,\vx_2) = ~&\chi^\free(\eta_1,\vx_1,\eta_2,\vx_2) + \frac{\lambda}{(4\pi)^3} \int_0^{\eta_1} d \eta_1' \int d^3 \vx_1' \int_0^{\eta_2} d \eta_2' \int d^3  \vx_2'\nonumber\\
&\times ~ a(\eta_1') a(\eta_2') \frac{\delta(\eta_1-\eta_1' - |\vx_1-\vx_1'|)}{|\vx_1-\vx_1'|}\frac{\delta(\eta_2-\eta_2' - |\vx_2-\vx_2'|)}{|\vx_2-\vx_2'|}\nonumber\\
&\times ~ \delta((\eta_1'-\eta_2')^2-|\vx_1'-\vx_2'|^2) \chi(\eta_1',\vx_1',\eta_2',\vx_2').
\label{eq:inteqcurvedexplicitchi}
\end{align}
We can see that this equation has almost exactly the same form as the massless version of \eqref{eq:inteq} on $\tfrac{1}{2}\M$ (see \eqref{eq:a0informal}). The only difference is the additional appearance of the factor $a(\eta_1') a(\eta_2')$ inside the integrals.

Going through the same steps as for \eqref{eq:defa0} before, \eqref{eq:inteqcurvedexplicitchi} can be defined on test functions $\chi \in \mathcal{S}$ by
\be
	\chi ~=~ \chi^\free + \widetilde{A}_0 \chi,
\label{eq:inteqcurvedabstract}
\ee
where $\widetilde{A}_0$ is defined by (using coordinates $x=(\eta_1,\vx)$, $y=(\eta_2,\vy)$):
\begin{align}\nonumber
    &(\widetilde{A}_0\psi)(x,y)=\frac{\lambda}{(4\pi)^3}\int_{B_{y^0}(\vy)}d^3\vy'  \int_0^{2\pi}d\varphi \int_{-1}^{1} d\!\cos\vartheta ~\frac{|b^2|}{4(b^0+|\vb|\cos\vartheta)^2 |\vy'|} ~\times \\
    &a(\eta_1+\eta_1') a(\eta_2+\eta_2') \psi(x+x',y+y') \left(1_{b^2>0}1_{b^0>0} 1_{\cos\vartheta > \frac{b^2}{2x^0|\vb|} - \frac{b^0}{|\vb|}}+1_{b^2<0}1_{\cos\vartheta<\frac{b^2}{2x^0|\vb|} - \frac{b^0}{|\vb|}}\right),
\label{eq:defa0tilde}
\end{align}
with \(\eta_1'=-r^*=-|\vx'| , \eta_2'=-|\vy'|\). (Here, $b$ and $r^*$ are defined as in \eqref{eq:b} and \eqref{eq:r}, respectively).

Knowing precisely how our integral equation on the flat FLRW spacetime is to be understood, we can formulate the respective existence and uniqueness theorem:

\begin{theorem}[Existence of dynamics for an open FLRW universe.]
	\label{thm:existencecurved}~\\
	Let $a: [0,\infty) \rightarrow [0,\infty)$ be a continuous function with $a(0)=0$ and $a(\eta)>0$ for $\eta>0$. Moreover, let
	\be
		g(t) = \exp \left( \gamma \int_0^t d\tau \, a(\tau) \right).
	\ee
	Then, the operator $\widetilde{A}_0$ satisfies the following estimate:
	\be
		\sup_{\chi \in \mathcal{S}\big( ([0,\infty)\times \R^3)^2\big)} \frac{\| \widetilde{A}_0 \chi \|_g}{\| \chi \|_g} ~\leq~ \frac{\lambda}{8\pi \gamma^2}.
	\label{eq:a0tildebound}
	\ee
	$\widetilde{A}_0$ can be extended to a linear operator on $\Banach_g$ which satisfies the same bound. Moreover, for $\gamma < \sqrt{\frac{\lambda}{8\pi}}$, the equation $\chi = \chi^\free + \widetilde{A}_0\chi$ has a unique solution $\chi \in \Banach_g$ for every $\psi^\free \in \Banach_g$.
\end{theorem}

The proof can be found in Sec.\@ \ref{sec:proofexistencecurved}.

\begin{remarks}
	\begin{enumerate}
		\item \textit{Manifest covariance.} The theorem shows the existence and uniqueness of solutions of the manifestly covariant integral equation \eqref{eq:inteqcurved}. Our example of a particular FLRW spacetime thus achieves its goal of demonstrating that a cutoff in time can arise naturally in a cosmological context.
		\item \textit{Initial value problem.} As in the case of $\tfrac{1}{2}\M$, the solution $\chi$ satisfies $\chi(0,\vx,0,\vy) = \chi^\free(0,\vx,0,\vy)$ where $\chi^\free$ is determined by the solution $\psi^\free$ of the free conformal wave equation \eqref{eq:conformalwaveeq} in both spacetime variables. Since $\psi^\free$ is determined by initial data at $\eta_1 = 0 =\eta_2$, so are $\chi^\free$ and $\chi$.
		\item \textit{Behavior of $\psi$ towards the Big Bang singularity}. While the transformed wave function $\chi$ remains bounded for $\eta_1, \eta_2 \rightarrow 0$, the physical wave function $\psi(\eta_1,\vx,\eta_2,\vy) = \frac{1}{a(\eta_1)a(\eta_2)} \chi(\eta_1,\vx,\eta_2,\vy)$ diverges like $\frac{1}{a(\eta_1)a(\eta_2)}$. This is to be expected, as the Klein-Gordon equation has a preserved "energy" (given by a certain spatial integral) and as the volume in $\vx, \vy$ contracts to zero towards the Big Bang.
		\item \textit{$N$-particle generalization.} As shown in Sec.\@ \ref{sec:Npart} for the Minkowski half-space, it would also be easy to extend Thm.\@ \ref{thm:existencecurved} to $N$ particles. To avoid duplication, we do not carry this out explicitly for the curved spacetime example here.
	\end{enumerate}
\end{remarks}

\section{Proofs}
\label{sec:proofs}

\subsection{Proof of Theorem \ref{thm:bounds}} \label{sec:proofbounds}
The proof is divided into the proofs of the estimates \eqref{eq:estimatea0}, \eqref{eq:estimatea1}, \eqref{eq:estimatea2} and \eqref{eq:estimatea12}, respectively. Here, \eqref{eq:estimatea0} is the most singular and difficult term which deserves the greatest attention.

Throughout this subsection, let $\psi \in \mathcal{S}((\tfrac{1}{2}\M)^2)$.

\subsubsection{Estimate of the massless term \eqref{eq:estimatea0}.} \label{sec:estimatemassless}

We start with Eq.\@ \eqref{eq:defa0} and take the absolute value. Using, in addition, that
\be
	|\psi(x,y)| ~\leq~ \| \psi \|_g \, g(x^0) g(y^0)
\ee
leads us to:
\begin{align}
    &|A_0\psi|(x,y) \le \frac{\lambda \|\psi\|_g}{4(4\pi)^3} \int_{B_{y^0}(\vy)}d^3 \vy' \int_0^{2\pi} d\varphi \int_{-1}^1 d\cos\vartheta \, \frac{|b^2|}{(b^0+|\vb|\cos\vartheta)^2 |\vy'|} g(y^0-|\vy'|)\nonumber\\
    &\times g\left(x^0-\frac{1}{2}\frac{b^2}{b^2+|\vb|\cos\vartheta}\right)
    \left(1_{b^2>0}1_{b^0>0} 1_{\cos\vartheta > \frac{b^2}{2x^0|\vb|} - \frac{b^0}{|\vb|}}+1_{b^2<0}1_{\cos\vartheta<\frac{b^2}{2x^0|\vb|} - \frac{b^0}{|\vb|}}\right).
\label{eq:a0calc01}
\end{align}
Next, we observe that the fraction $\frac{|b^2|}{(b^0+|\vb|\cos\vartheta)^2 |\vy'|}$ is the derivative of the fraction which occurs in the argument of the second $g$-function. Introducing $u = \cos \vartheta$ allows us to rewrite \eqref{eq:a0calc01} as
\begin{align}
     \eqref{eq:a0calc01} &=\frac{\lambda \|\psi\|_g}{8(4\pi)^2} \int_{B_{y^0}(\vy)}d^3\vy' \int_{-1}^1 du ~ 2 \sgn(b^2)\, \partial_u g_1\left(x^0-\frac{1}{2}\frac{b^2}{b^0+|\vb|u}\right) g(y^0-|\vy'|)\\\nonumber
    &~~~\times \left(1_{b^2>0}1_{b^0>0} 1_{u > \frac{b^2}{2x^0|\vb|} - \frac{b^0}{|\vb|}}+1_{b^2<0}1_{u<\frac{b^2}{2x^0|\vb|} - \frac{b^0}{|\vb|}}\right) \frac{1}{|\vb||\vy'|}\\\label{uIntegral}
    &=\frac{\lambda \|\psi\|_g}{4(4\pi)^2} \int_{B_{y^0}(\vy)}d^3\vy' \int_{-1}^1 du  ~ \partial_u g_1\left(x^0-\frac{1}{2}\frac{b^2}{b^0+|\vb|u}\right) g(y^0-|\vy'|)\\\nonumber
    &~~~\times   \left(1_{b^2>0}1_{b^0>0} 1_{u > \frac{b^2}{2x^0|\vb|} - \frac{b^0}{|\vb|}}-1_{b^2<0}1_{u<\frac{b^2}{2x^0|\vb|} - \frac{b^0}{|\vb|}}\right) \frac{1}{|\vb||\vy'|}. 
\label{eq:a0calc02}
\end{align}
This form allows for a direct integration with respect to $u$.
Before we integrate, we check whether the conditions implicit in the characteristic functions can always be satisfied. (Otherwise, the respective term would not contribute any further and we could drop it.) Recall that $	b = x-y-(-|\vy'|, \vy')$. First we check whether in the case \(b^2>0, b^0>0\) it is true that \( 1> \frac{b^2}{2x^0|\vb|} - \frac{b^0}{|\vb|}\) holds. (The comparison with 1 is due to the upper range for $u$.) We compute
\begin{align}
    1> \frac{b^2}{2x^0|\vb|} - \frac{b^0}{|\vb|} &\iff  2 x^0 |\vb|+2 x^0 b^0 > b^2 \nonumber\\
    &\iff 2 x^0(b^0 + |\vb|) > (b^0+|\vb|)(b^0-|\vb|)\nonumber\\
   &\!\!\!\! \overset{b^2>0,b^0>0}{\iff} 2x^0>b^0-|\vb| \nonumber \\
   &\iff  x^0+y^0 - |\vy'| > -|\vb|.
\end{align}
Now because of \(|\vy'|<y^0\) we see that this inequality always holds true. Hence the respective term in \eqref{eq:a0calc02} contributes without further restrictions.

Next, we turn to the case \(b^2<0\). Here we check whether (or when) $-1<\frac{b^2}{2x^0|\vb|} - \frac{b^0}{|\vb|}$ holds. (The comparison with $-1$ is due to the lower bound for $u$.) A similar calculation yields 
\begin{align}
    -1<\frac{b^2}{2x^0|\vb|} - \frac{b^0}{|\vb|} &\iff -2x^0|\vb| + 2x^0 |\vb| < b^2\\
    &\iff 2x^0 (b^0-|\vb|) < (b^0-|\vb|)(b^0+|\vb|) \\
   & \overset{b^2<0}{\iff} 2x^0 > b^0 + |\vb|.
\end{align}
This inequality need not always hold, as we can increase \(|\vb|\) with respect to \(b^0\) as much as we like, e.g., by picking \(|\vx-\vy|\) large. Therefore, in this case, the respective term is only sometimes nonzero. We make this clear by including the characteristic function $1_{2x^0>b^0+|\vb|}$.

Taking these considerations into account, we now carry out the $u$-integration in \eqref{eq:a0calc02}:
\begin{align}
    &|A_0\psi|(x,y) ~\le~ \frac{\lambda \|\psi\|_g}{4(4\pi)^2} \int_{B_{y^0}(\vy)}d^3\vy'~ \frac{g(y^0-|\vy'|)}{|\vb||\vy'|} \\
    &\times \left(1_{b^2>0,b^0>0} \left[g_1\left(x^0-\frac{1}{2} \frac{b^2}{b^0+|\vb|}\right) - g_1\left( x^0 - \frac{1}{2}\frac{b^2}{b^0+|\vb| \max (-1, \frac{b^2}{2x^0|\vb|} - \frac{b^0}{|\vb|})}\right)\right]\right.\\
    &\left. -1_{b^2<0}1_{2x^0>b^0+|\vb|}\left[
    g_1\left(x^0-\frac{1}{2} \frac{b^2}{b^0+|\vb| \min(1, \frac{b^2}{2x^0|\vb|}-\frac{b^0}{|\vb|})}\right)- g_1\left(x^0-\frac{1}{2} \frac{b^2}{b^0-|\vb|} \right)\right]\right).
\label{eq:a0calc03}
\end{align}
The minima and maxima in this expression result from the indicator functions $1_{u > \frac{b^2}{2x^0|\vb|} - \frac{b^0}{|\vb|}}$ and $1_{u<\frac{b^2}{2x^0|\vb|} - \frac{b^0}{|\vb|}}$, respectively.

Our next step is to simplify the complicated fractions in \eqref{eq:a0calc03} involving $\min$ and $\max$. For the first one we use that \(1/\max(a,b)=\min(1/a,1/b)\) whenever \(a,b>0\) or \(a,b<0\) holds. Therefore, we have:
\begin{align}
    \frac{1}{2}\frac{b^2}{b^0+|\vb|\max\left( -1,\frac{b^2}{2x^0|\vb|} - \frac{b^0}{|\vb|}\right)}
	&= \frac{1}{2} \frac{b^2}{\max\left(b^0-|\vb|,\frac{b^2}{2x^0}\right)}\nonumber\\
    &= \frac{1}{2} \min \left(\frac{b^2}{b^0-|\vb|}, 2x^0 \right)\nonumber\\
    &= \min \left( \frac{b^0+|\vb|}{2}, x^0 \right).
\end{align}
The fraction in \eqref{eq:a0calc03} which contains a minimum can be simplified by observing that
\begin{align}
    b^0+|\vb|\min\left(1,\frac{b^2}{2x^0|\vb|}-\frac{b^0}{|\vb|}\right) = \min \left( b^0+|\vb|, \frac{b^2}{2x^0} \right)= \frac{b^2}{2x^0}
\end{align}
as the term contributes only for \(b^2<0\) and $2x^0>b^0+|\vb|$ (note that then $\frac{b^2}{2x^0} < \frac{(b^0)^2-|\vb|^2}{b^0 + |\vb|} = b^0 - |\vb| < b^0 + |\vb|$). Thus,
\be
	\frac{1}{2} \frac{b^2}{b^0+|\vb| \min(1, \frac{b^2}{2x^0|\vb|}-\frac{b^0}{|\vb|})} = x^0.
\ee
With these simplifications, we obtain (using $g_1(0)=0$):
\begin{align}
    &|A_0\psi|(x,y)~ \le ~ \frac{\lambda \|\psi\|_g}{4(4\pi)^2} \int_{B_{y^0}(\vy)}d^3\vy'~ \frac{g(y^0-|\vy'|)}{|\vb||\vy'|} \nonumber\\
    &~~~\times \left(1_{b^2>0,b^0>0} \left[g_1\left(x^0-\frac{b^0-|\vb|}{2} \right) - g_1\left( x^0 - \min\left(\frac{b^0+|\vb|}{2},x^0\right)\right)\right]\right.\nonumber\\
    &\left. ~~~-1_{b^2<0}~1_{2x^0>b^0+|\vb|}\left[
    g_1\left(x^0-x^0\right)- g_1\left(x^0-\frac{b^0+|\vb|}{2} \right)\right]\right)\nonumber\\\label{y'Integral1}
    &=\frac{\lambda \|\psi\|_g}{4(4\pi)^2} \int_{B_{y^0}(\vy)}d^3\vy' \frac{g(y^0-|\vy'|)}{|\vb||\vy'|}
    1_{b^2>0,b^0>0} \, g_1\left(\frac{x^0+y^0-|\vy'|+|\vb|}{2} \right) \\\label{y'Integral2}
    &~~~-\frac{\lambda \|\psi\|_g}{4(4\pi)^2} \int_{B_{y^0}(\vy)}d^3\vy' \, \frac{g(y^0-|\vy'|)}{|\vb||\vy'|}
    1_{b^2>0,b^0>0}\,  g_1\left( \max\left(\frac{x^0+y^0-|\vy'|-|\vb|}{2},0\right)\right)\\\label{y'Integral3}
    &~~~ +\frac{\lambda \|\psi\|_g}{4(4\pi)^2} \int_{B_{y^0}(\vy)}d^3\vy' \, \frac{g(y^0-|\vy'|)}{|\vb||\vy'|} 1_{b^2<0}~1_{x^0+y^0-|\vy'|>|\vb|}
     ~g_1\left(\frac{x^0+y^0-|\vy'|-|\vb|}{2} \right).
\end{align}

We now want to carry out as many of the remaining $\vy'$-integrations as possible. In order to do so, we orient the coordinates such that \(\vx-\vy\) is parallel to the \((\vy')_3\) axis. Then the integrands in \eqref{y'Integral1}-\eqref{y'Integral3} are independent of the azimuthal angle $\varphi$ of the respective spherical coordinate system $(\rho,\theta,\varphi)$ with standard conventions.

In order to perform the remaining angular and then the radial integral, we need to find out which boundaries for $\theta$ and $r$ result from the characteristic functions. First we analyze for which arguments the maximum in \eqref{y'Integral2} is greater than zero and therefore contributes to the integral (as $g_1(0) = 0$). We have:
\begin{align}
    \frac{x^0+y^0-|\vy'|-|\vb|}{2}~>~0 &\iff  (x^0+y^0-|\vy'|)^2>|\vx-\vy|^2+|\vy'|^2+2|\vy'||\vx-\vy|\cos\theta\nonumber\\
   &\iff \cos\theta<\frac{(x^0+y^0)^2}{2|\vy'||\vx-\vy|} -\frac{|\vx-\vy|}{2|\vy'|} - \frac{x^0+y^0}{|\vx-\vy|}=:P_{x,y}(|\vy'|).
\label{eq:costhetapxy}
\end{align}
This calculation also helps to reformulate the second indicator function $1_{b^2<0}~1_{x^0+y^0-|\vy'|>|\vb|}$ in \eqref{y'Integral3} (for which we have \(b^2<0\)).
The condition \(b^0>0\) in \eqref{y'Integral1} and \eqref{y'Integral2} is readily seen to be equivalent to
\be
	|\vy'|>y^0-x^0.
	\label{eq:b0cond}
\ee
In order to perform the \(\theta\)-integral we have to translate \(b^2\gtrless 0\) into conditions on \(\theta\). We have:
\begin{align}
    b^2>0 &\iff (x^0-y^0+|\vy'|)^2~>~ |\vx-\vy|^2+|\vy'|^2+2|\vy'||\vx-\vy|\cos\theta \nonumber\\
    &\iff \cos\theta~<~ \frac{(x-y)^2}{2|\vy'||\vx-\vy|} + \frac{x^0-y^0}{|\vx-\vy|}:=K_{x-y}(|\vy'|).
\label{eq:costhetakxy}
\end{align}
With these considerations, we have extracted relatively simple conditions on the boundaries of the integrals in spherical coordinates. However, if different restrictions of the boundaries conflict with each other, it may happen that for some parameter values the domain of integration is the empty set. 
We check whether this is so term by term, focusing on the \(\theta\)-integration first. For term \eqref{y'Integral1}, \(\theta\) needs to satisfy \(-1<\cos\theta<\min(1,K_{x-y}(|\vy'|))\), so we need to check whether \(-1<K_{x-y}(|\vy'|)\) holds. We have:
\begin{align}\label{-1<K}
    -1 < K_{x-y}(|\vy'|) &\iff -2|\vy'||\vx-\vy|<(x-y)^2+2|\vy'|(x^0-y^0)\nonumber\\
    &\iff 0~<~(x-y)^2+2|\vy'|(x^0-y^0+|\vx-\vy|)\nonumber\\
    &\iff \left\{\begin{matrix}
    \frac{y^0-x^0+|\vx-\vy|}{2}<|\vy'|  \quad \text{for } |\vx-\vy|>y^0-x^0\\
    \frac{y^0-x^0+|\vx-\vy|}{2}>|\vy'| \quad \, \text{for } |\vx-\vy|<y^0-x^0.
    \end{matrix}\right.
\end{align}
Together with \eqref{eq:b0cond}, we obtain the condition \(y^0-x^0<|\vy'|<\frac{y^0-x^0+|\vx-\vy|}{2}<y^0-x^0\) in the second case which means that there is no contribution to the integral. So we focus on the first case,
\be
	\frac{y^0-x^0+|\vx-\vy|}{2}<|\vy'|  \quad \text{and } |\vx-\vy|>y^0-x^0,
\label{eq:xycond1}
\ee
by including the characteristic function $1_{|\vx-\vy|>y^0-x^0}$ in the integral. Next, we turn to the radial integral. By comparing its upper limit $|\vy'|<y^0$ and lower limit $(y^0-x^0+|\vx-\vy|)/2$, we find that the integral can only be nonzero for
\be
	y^0+x^0>|\vx-\vy|.
	\label{eq:xycond2}
\ee
We make this clear by including the respective characteristic function.

\paragraph{Simplification of term \eqref{y'Integral1}.}
These considerations allow us to continue computing \eqref{y'Integral1}:
\begin{align}\label{casesFirstTermBeginning}
    \eqref{y'Integral1}
    &=\frac{\lambda \|\psi\|_g}{4(4\pi)^2}1_{y^0+x^0>|\vx-\vy|}  \int_{\max(0,y^0-x^0)}^{y^0} d\rho \int_{0}^{2\pi}d\varphi~ 
    1_{\frac{y^0-x^0+|\vx-\vy|}{2}<\rho}1_{|\vx-\vy|>y^0-x^0}\nonumber\\ 
    &~~~\times\int_{-1}^{\min(1,K_{x-y}(\rho))} d\cos\theta  ~\frac{\rho \, g(y^0-\rho)}{\sqrt{|\vx-\vy|^2+\rho^2 + 2 |\vx-\vy|\rho \cos\theta}}\nonumber\\ 
	&~~~\times g_1\left(\frac{x^0+y^0-\rho + \sqrt{|\vx-\vy|^2+\rho^2+2\rho|\vx-\vy|\cos\theta}}{2}\right).
    \end{align}
Now we carry out the $\varphi$-integration and use the same trick for the $\theta$-integral as for the $\vartheta$-integral in the \(\vx'\)-integration earlier.  Moreover, we absorb some of the restrictions of \(\rho\) into the limits of the integrals. This yields:
    \begin{align}
    \eqref{y'Integral1} &=\frac{\lambda \|\psi\|_g}{8(4\pi)}  1_{y^0+x^0>|\vx-\vy|>y^0-x^0} \int_{\max\left(0,y^0-x^0,\frac{y^0-x^0+|\vx-\vy|}{2}\right)}^{y^0} d\rho \int_{-1}^{\min(1,K_{x-y}(\rho))} \!\!\! dw \, \frac{2 g(y^0-\rho)}{|\vx-\vy|}\nonumber\\
 &~~~\times \partial_w g_2\left(\frac{x^0+y^0-\rho + \sqrt{|\vx-\vy|^2+\rho^2+2\rho|\vx-\vy|w}}{2}\right)\nonumber\\
    &= \frac{\lambda \|\psi\|_g}{4(4\pi)} 1_{x^0+y^0>|\vx-\vy|>y^0-x^0} \int_{\max\left(0,y^0-x^0,\frac{y^0-x^0+|\vx-\vy|}{2}\right)}^{y^0} d\rho ~  \frac{ g(y^0-\rho)}{|\vx-\vy|}\nonumber\\ 
    &~~~\times \left[ g_2\left(\frac{x^0+y^0-\rho + \sqrt{|\vx-\vy|^2+\rho^2+2\rho|\vx-\vy|\min(1,K_{x-y}(\rho))}}{2}\right)\right.\nonumber\\
    &~~~\left.-g_2\left(\frac{x^0+y^0-\rho + ||\vx-\vy|-\rho|}{2}\right)\right]
\end{align}
The square root can be simplified using the following identity:
\begin{align}\label{PlugKIn}
   \sqrt{|\vx-\vy|^2+\rho^2+2\rho|\vx-\vy|K_{x-y}(\rho)}=
   \sqrt{\rho^2 +(x^0-y^0)^2+2\rho (x^0-y^0)}
   =|x^0-y^0+\rho|.
\end{align}
Using this, we can effectively pull the minimum out of the square root. We obtain:
\begin{align}\nonumber
    &\eqref{y'Integral1} =\frac{\lambda \|\psi\|_g}{16\pi} 1_{x^0+y^0>|\vx-\vy|>y^0-x^0} \int_{\max\left(0,y^0-x^0,\frac{y^0-x^0+|\vx-\vy|}{2}\right)}^{y^0} d\rho ~ \frac{ g(y^0-\rho)}{|\vx-\vy|}\\ 
    &\times\left[ g_2\left(\frac{x^0+y^0-\rho + \min(|\vx-\vy|+\rho,|x^0-y^0+\rho|)}{2}\right)-g_2\left(\frac{x^0+y^0-\rho + ||\vx-\vy|-\rho|}{2}\right)\right].
\label{CasesFirstTermEnd}
\end{align}
Next, we subdivide the conditions in the first indicator function into two cases, (a) $(x-y)^2>0$ and (b) $(x-y)^2<0$. In case (a), the condition $|\vx-\vy|>y^0-x^0$ implies $x^0>y^0$. This, in turn, yields $\max\left(0,y^0-x^0,\frac{y^0-x^0+|\vx-\vy|}{2}\right) = 0$.  Moreover, the condition $x^0+y^0>|\vx-\vy|$ is automatically satisfied (note that $x^0,y^0>0$). In case (b), the condition $|\vx-\vy|>0$ is automatically satisfied. We find:
\begin{align}
 \nonumber
     &\eqref{y'Integral1} = \frac{\lambda \|\psi\|_g}{16\pi} 1_{(x-y)^2>0, x^0>y^0} \int_{0}^{y^0} d\rho  \, \frac{ g(y^0-\rho)}{|\vx-\vy|}\\\nonumber
    &~~~\times\left[ g_2\left(\frac{x^0+y^0 + |\vx-\vy|}{2}\right)
    -g_2\left(\frac{x^0+y^0-\rho + ||\vx-\vy|-\rho|}{2}\right)\right]\\\nonumber
    &~~~+ \frac{\lambda \|\psi\|_g}{16\pi} 1_{(x-y)^2<0}~1_{ x^0+y^0>|\vx-\vy|} \int_{\frac{y^0-x^0+|\vx-\vy|}{2}}^{y^0} d\rho \, \frac{ g(y^0-\rho)}{|\vx-\vy|}\\\nonumber
    &~~~\times \left[ g_2\left(\frac{x^0+y^0-\rho + |x^0-y^0+\rho|}{2}\right)
    -g_2\left(\frac{x^0+y^0-\rho + ||\vx-\vy|-\rho|}{2}\right)\right]\\\nonumber
    &= \frac{\lambda \|\psi\|_g}{16\pi} 1_{(x-y)^2>0, x^0>y^0} \int_{0}^{y^0} d\rho \, \frac{ g(y^0-\rho)}{|\vx-\vy|}\\\nonumber
    &~~~\times \left[ g_2\left(\frac{x^0+y^0 + |\vx-\vy|}{2}\right)
    -g_2\max\left(\frac{x^0+y^0- |\vx-\vy|}{2},\frac{x^0+y^0+ |\vx-\vy|}{2}-\rho\right)\right]\\\nonumber
    &~~~+ \frac{\lambda \|\psi\|_g}{16\pi} 1_{(x-y)^2<0}~1_{ x^0+y^0>|\vx-\vy|} \int_{\frac{y^0-x^0+|\vx-\vy|}{2}}^{y^0} d\rho \, \frac{ g(y^0-\rho)}{|\vx-\vy|}\\
    &~~~\times\left[ g_2\max\left(x^0,y^0-\rho\right)
    -g_2\max\left(\frac{x^0+y^0-|\vx-\vy|}{2},\frac{x^0+y^0+|\vx-\vy|}{2}-\rho\right)\right].
\label{eq:resulty'Integral1}
\end{align}
Here and in the following we abbreviate $g_2(\max(\cdots))$ as $g_2 \max (\cdots)$, and similarly for the minimum. This ends the calculation of \eqref{y'Integral1}: we have arrived at an expression where no more exact calculations can be done and further estimates are needed.

\paragraph{Simplification of term \eqref{y'Integral2}.}
Next, we proceed with \eqref{y'Integral2} in a similar fashion. In case the reader is not interested in the details of the calculation, the result can be found in \eqref{eq:resulty'Integral2}.

The restrictions of the integration variables for \eqref{y'Integral2} are the same as for \eqref{y'Integral1}, namely:
\begin{align}
\cos\theta<K_{x-y}(|\vy'|)\quad\quad   &\text{from } \eqref{eq:costhetakxy},\\
\frac{y^0-x^0+|\vx-\vy|}{2} < |\vy'| \quad\quad  &\text{from } \eqref{eq:xycond1}\label{eq:condfromkxy}\\
y^0-x^0<|\vx-\vy|<y^0+x^0 \quad\quad  &\text{from \eqref{eq:xycond1} and from } \eqref{eq:xycond2}.
\end{align}
The only difference is that from the maximum in \eqref{y'Integral2}, we obtain the additional restriction \eqref{eq:costhetapxy}, i.e.
\be
	\cos\theta< P_{x,y}(|\vy'|).
\ee
We need to check if there are new restrictions imposed by \(P_{x,y}(|\vy'|)>-1\). We compute
\begin{align}
    P_{x,y}(|\vy'|) ~&>~ -1~~ \iff\nonumber\\
     \frac{(x^0+y^0)^2}{2|\vy'||\vx-\vy|} - \frac{|\vx-\vy|}{2|\vy'|} - \frac{x^0+y^0}{|\vx-\vy|}~&>~-1    ~~\iff\nonumber \\
   |\vy'| ~&<~ \frac{x^0+y^0+|\vx-\vy|}{2};
\end{align}
however, the last inequality is already ensured by \eqref{eq:condfromkxy} and $x^0>0$. 
In order to be able to evaluate \eqref{y'Integral2} further, we next plug the condition $\cos \theta < P_{x,y}(|\vy'|)$ into the expression for $|\vb|$. This yields (recall that we use spherical variables for $|\vy'|$):
\begin{align}\label{PlugPIn}
   |\vb| &= \sqrt{|\vx-\vy|^2+\rho^2 + 2 \rho |\vx-\vy| \cos \theta} < \sqrt{|\vx-\vy|^2+\rho^2 + 2 \rho |\vx-\vy| P_{x,y}(\rho)}\nonumber\\
    &= \sqrt{\rho^2 - 2\rho(x^0+y^0) +(x^0+y^0)^2}=x^0+y^0-\rho.
\end{align}
With this, we perform for \eqref{y'Integral2} the analogous calculation to \eqref{casesFirstTermBeginning}--\eqref{CasesFirstTermEnd}. This yields:
\begin{align}
    \eqref{y'Integral2}&=\frac{\lambda\|\psi\|_g}{16\pi} 1_{y^0-x^0<|\vx-\vy|<x^0+y^0} \int_{\max \left(0,y^0-x^0, \frac{y^0-x^0+|\vx-\vy|}{2} \right)}^{y^0} d\rho ~\frac{g(y^0-\rho)}{|\vx-\vy|}\nonumber\\
    &~~~\times \left[g_2\left(\frac{x^0+y^0-\rho-\min(|\vx-\vy|+\rho,|x^0-y^0+\rho|,x^0+y^0-\rho)}{2} \right)\right.\nonumber\\
    &~~~\left.-g_2\left( \frac{x^0+y^0-\rho-||\vx-\vy|-\rho|}{2} \right) \right] \nonumber\\
    &= \frac{\lambda \|\psi\|_g}{16\pi} 1_{(x-y)^2>0, x^0>y^0} \int_{0}^{y^0} d\rho  ~\frac{ g(y^0-\rho)}{|\vx-\vy|}\nonumber\\
    &~~~\times\left[ g_2\left(\frac{x^0+y^0 - |\vx-\vy|}{2}-\rho\right)
    -g_2\left(\frac{x^0+y^0-\rho - ||\vx-\vy|-\rho|}{2}\right)\right]\nonumber\\
    &~~~+ \frac{\lambda \|\psi\|_g}{16\pi} 1_{(x-y)^2<0}~1_{ x^0+y^0>|\vx-\vy|} \int_{\frac{y^0-x^0+|\vx-\vy|}{2}}^{y^0} d\rho  ~\frac{ g(y^0-\rho)}{|\vx-\vy|}\nonumber\\
    &~~~\times \left[ g_2\left(\frac{x^0+y^0-\rho - |x^0-y^0+\rho|}{2}\right)
    -g_2\left(\frac{x^0+y^0-\rho - ||\vx-\vy|-\rho|}{2}\right)\right]\nonumber\\
    &= \frac{\lambda \|\psi\|_g}{16\pi} 1_{(x-y)^2>0, x^0>y^0} \int_{0}^{y^0} d\rho  ~\frac{ g(y^0-\rho)}{|\vx-\vy|}\nonumber\\
    &~~~\times \left[ g_2\left(\frac{x^0+y^0 - |\vx-\vy|}{2}-\rho\right)
    -g_2\min\left(\frac{x^0+y^0 - |\vx-\vy|}{2},\frac{x^0+y^0 + |\vx-\vy|}{2}-\rho\right)\right]\nonumber\\
    &~~~+ \frac{\lambda \|\psi\|_g}{16\pi} 1_{(x-y)^2<0}~1_{ x^0+y^0>|\vx-\vy|} \int_{\frac{y^0-x^0+|\vx-\vy|}{2}}^{y^0} d\rho ~ \frac{ g(y^0-\rho)}{|\vx-\vy|}\nonumber\\
    &~~~\times \left[ g_2\min\left(x^0,y^0-\rho\right)
    -g_2\min\left(\frac{x^0+y^0- |\vx-\vy|}{2},\frac{x^0+y^0+ |\vx-\vy|}{2}-\rho\right)\right].
\label{eq:resulty'Integral2}
\end{align}
This ends the calculation of \eqref{y'Integral2}. 

\paragraph{Simplification of term \eqref{y'Integral3}.}
We next turn to \eqref{y'Integral3}. In case the reader is not interested in the details of the computation, the result can be found in \eqref{eq:resulty'Integral3}.
First we note that the restriction imposed by the first indicator function here is \(\cos\theta>K_{x-y}(|\vy'|)\) and the condition of the second indicator function is \(\cos\theta<P_{x,y}(|\vy'|)\). In order to to satisfy these conditions (and the restrictions of the regular range of integration) it is required that
\begin{equation}
    \max(-1,K_{x-y}(|\vy'|)<\cos\theta<\min(1,P_{x,y}(|\vy'|)).
\end{equation}
This leads us to ask which restrictions on \(|\vy'|\) are imposed by the conditions
\begin{align}
    K_{x-y}(|\vy'|) ~&<~1,\\
    P_{x,y}(|\vy'|) ~&>~-1,\\
    K_{x-y}(|\vy'|) ~&<~ P_{x,y}(|\vy'|).
\end{align}
These restrictions shall be computed next. With \(|\vy'|=\rho\), we find:
\begin{align}
     K_{x-y}(|\vy'|)<1~~~
    &\iff~~~ \frac{(x-y)^2}{2\rho|\vx-\vy|}+\frac{x^0-y^0}{|\vx-\vy|}~<~1\nonumber\\
 &\iff~~~    (x-y)^2~<~2\rho(y^0-x^0+|\vx-\vy|)\nonumber\\
   &\iff~~~  \left\{\begin{matrix}\rho>\frac{y^0-x^0-|\vx-\vy|}{2}\quad \text{for } |\vx-\vy|>x^0-y^0,\\
\rho<\frac{y^0-x^0-|\vx-\vy|}{2}\quad \, \text{for } |\vx-\vy|<x^0-y^0.
     \end{matrix} \right.
\end{align}
The second case in the last line is in conflict with \(\rho>0\), so we have to impose the first condition on \eqref{y'Integral3}. We continue with \(P_{x,y}(\rho)>-1\).
\begin{align}
    P_{x,y}(\rho)>-1~&\iff~ \frac{(x^0+y^0)^2}{2\rho|\vx-\vy|} - \frac{|\vx-\vy|}{2\rho}-\frac{x^0+y^0}{|\vx-\vy|} ~>~-1 \nonumber \\
   &\iff~ (x^0+y^0)^2-|\vx-\vy|^2~>~2\rho(x^0+y^0-|\vx-\vy|)\nonumber \\
    &\iff~\left\{\begin{matrix}\rho< \frac{x^0+y^0+|\vx-\vy|}{2} \quad \text{for } x^0+y^0>|\vx-\vy|,\\
    \rho> \frac{x^0+y^0+|\vx-\vy|}{2} \quad \text{for } x^0+y^0<|\vx-\vy|.
    \end{matrix} \right.
\end{align}
The second case is in conflict with \(\rho<y^0\), so we implement indicator functions corresponding only to the first case in \eqref{y'Integral3}. The third condition \(K_{x-y}(\rho)<P_{x,y}(\rho)\) in fact does not impose any additional conditions. This can be seen as follows:
\begin{align}
    K_{x-y}(\rho)~<~P_{x,y}(\rho)~~~&\iff~~~ \frac{(x-y)^2}{2\rho|\vx-\vy|}+\frac{x^0-y^0}{|\vx-\vy|}~<~\frac{(x^0+y^0)^2}{2\rho|\vx-\vy|} - \frac{|\vx-\vy|}{2\rho}-\frac{x^0+y^0}{|\vx-\vy|}\nonumber\\
   &\iff~~~ -2x^0y^0 + 4\rho x^0~<~2x^0y^0\nonumber\\
   &\iff~~~   \rho<y^0,
\end{align}
which always holds true.

Taking into account the computed restrictions, we arrive at:
\begin{align}
    &\eqref{y'Integral3} \stackrel{\cos\theta=w}{=}\frac{\lambda \|\psi\|_g}{4(4\pi)^2} \int_0^{2\pi} d\varphi \int_0^{y^0} d\rho \int_{-1}^1 dw \,1_{K_{x-y}(\rho)<w<P_{x,y}(\rho)} 1_{\frac{y^0-x^0-|\vx-\vy|}{2}<\rho<\frac{x^0+y^0+|\vx-\vy|}{2}} \nonumber\\
     &~~~\times 1_{x^0-y^0<|\vx-\vy|<x^0+y^0} \frac{g(y^0-\rho)\rho}{\sqrt{\rho^2+|\vx-\vy|^2+2\rho|\vx-\vy|w}}\nonumber\\
     &~~~\times g_1\left( \frac{x^0+y^0-\sqrt{\rho^2+|\vx-\vy|^2+2\rho|\vx-\vy|w}}{2}\right)\nonumber\\
     &= \frac{\lambda \|\psi\|_g 2\pi}{4(4\pi)^2}1_{x^0-y^0<|\vx-\vy|<x^0+y^0} \int_{\max\left(0,\frac{y^0-x^0-|\vx-\vy|}{2}\right)}^{\min\left(y^0,\frac{x^0+y^0+|\vx-\vy|}{2} \right)}d\rho
     \int_{\max(-1,K_{x-y}(\rho))}^{\min(1,P_{x,y}(\rho))}dw \, \frac{-2 g(y^0-\rho)}{|\vx-\vy|}\nonumber\\
     &~~~\times \partial_w 
     g_2\left(\frac{x^0+y^0-\sqrt{\rho^2+|\vx-\vy|^2 +2\rho|\vx-\vy|w}}{2} \right)\nonumber\\
     &=\frac{\lambda \|\psi\|_g}{16\pi} 1_{x^0-y^0<|\vx-\vy|<x^0+y^0} 
     \int_{\max\left(0,\frac{y^0-x^0-|\vx-\vy|}{2}\right)}^{\min\left(y^0,\frac{x^0+y^0+|\vx-\vy|}{2} \right)}d\rho \, \frac{g(y^0-\rho)}{|\vx-\vy|}\nonumber\\ 
     &~~~\times \left[g_2\left(\frac{x^0+y^0-\rho-\sqrt{\rho^2+|\vx-\vy|^2+2\rho|\vx-\vy|\max(-1,K_{x-y}(\rho))}}{2}\right)\right.\nonumber\\
     &~~~~\left.-g_2\left(\frac{x^0+y^0-\rho-\sqrt{\rho^2+|\vx-\vy|^2+2\rho|\vx-\vy|\min(1,P_{x,y}(\rho))}}{2}\right)
     \right].
\end{align}
At this point, the expressions look quite formidable. We can, however, achieve significant simplifications by inserting the functional form of $K_{x,y}(\rho)$ and $P_{x,y}(\rho)$ as in \eqref{PlugPIn} and \eqref{PlugKIn}. This yields:
\begin{align}
    \eqref{y'Integral3} &=
    \frac{\lambda\|\psi\|_g}{16\pi} 1_{x^0-y^0<|\vx-\vy|<x^0+y^0}
    \int_{\max\left(0,\frac{y^0-x^0-|\vx-\vy|}{2}\right)}^{\min\left(y^0,\frac{x^0+y^0+|\vx-\vy|}{2} \right)}d\rho \, \frac{g(y^0-\rho)}{|\vx-\vy|}\nonumber\\ 
    &~~~\times\left[ g_2\left( \frac{x^0+y^0-\rho-\max(||\vx-\vy|-\rho|,|x^0-y^0+\rho|)}{2} \right)\right.\nonumber\\
    &~~~\left.-g_2\left( \frac{x^0+y^0-\rho-\min(|\vx-\vy|+\rho,x^0+y^0-\rho)}{2} \right)\right]
	\label{eq:y'Integral3calc}
\end{align}
Now we simplify the arguments of the $g_2$-functions. For the first one, we have:
\begin{align}
	&x^0+y^0-\rho-\max(||\vx-\vy|-\rho|,|x^0-y^0+\rho|)\nonumber\\
&~~~=~ x^0+y^0-\rho -\max ( |\vx-\vy|-\rho,\rho-|\vx-\vy|,x^0-y^0+\rho,y^0-\rho-x^0)\nonumber\\
&~~~= \min\left(x^0+y^0-\rho-|\vx-\vy|,x^0+y^0+|\vx-\vy|-2\rho,2(y^0-\rho),2x^0 \right).
\end{align}
And for the second one:
\be
	x^0+y^0-\min(|\vx-\vy|+\rho,x^0+y^0-\rho) ~=~ \max(x^0+y^0-|\vx-\vy|-2\rho,0).
\ee
Using this in \eqref{eq:y'Integral3calc}, we find:
\begin{align}
    \eqref{y'Integral3} &=\frac{\lambda\|\psi\|_g}{16\pi} 
    1_{x^0-y^0<|\vx-\vy|<x^0+y^0}
    \int_{\max\left(0,\frac{y^0-x^0-|\vx-\vy|}{2}\right)}^{\min\left(y^0,\frac{x^0+y^0+|\vx-\vy|}{2} \right)}d\rho \, \frac{g(y^0-\rho)}{|\vx-\vy|}\nonumber\\
    &~~~\times \left[g_2 \min\left(\frac{x^0+y^0-|\vx-\vy|}{2},\frac{x^0+y^0+|\vx-\vy|}{2}-\rho,y^0-\rho,x^0 \right) \right.\nonumber\\
    &~~~\left. -g_2 \max\left( \frac{x^0+y^0-|\vx-\vy|}{2}-\rho,0\right) \right].
\end{align}
As in the consideration below \eqref{CasesFirstTermEnd}, we split the expression into separate terms with \((x-y)^2 \gtrless 0\). Using \(y^0\gtrless x^0+|\vx-\vy|\), we can simplify the expressions involving the minimum. This results in:
\begin{align}
    \eqref{y'Integral3} &=
    \frac{\lambda\|\psi\|_g}{16\pi} 1_{(x-y)^2>0,y^0>x^0}
    \int_{\frac{y^0-x^0-|\vx-\vy|}{2}}^{\frac{x^0+y^0+|\vx-\vy|}{2}}d\rho \, \frac{g(y^0-\rho)}{|\vx-\vy|}\nonumber\\
    &\times \left[g_2 \min\left(\frac{x^0+y^0+|\vx-\vy|}{2}-\rho,x^0\right) -g_2\max\left( \frac{x^0+y^0-|\vx-\vy|}{2}-\rho,0\right) \right]\nonumber\\
    &+\frac{\lambda\|\psi\|_g}{16\pi} 
    1_{(x-y)^2<0,|\vx-\vy|<x^0+y^0}
    \int_{0}^{y^0}d\rho \, \frac{g(y^0-\rho)}{|\vx-\vy|}\nonumber\\
    &\times \left[g_2\min\left(\frac{x^0+y^0-|\vx-\vy|}{2},y^0-\rho\right) -g_2\max\left( \frac{x^0+y^0-|\vx-\vy|}{2}-\rho,0\right) \right].
\label{eq:resulty'Integral3}
\end{align}
This concludes the calculation of \eqref{y'Integral3}.

\paragraph{Summary of the first estimate.} We have obtained the following bound for $| A_0 \psi |(x,y)$:
\begin{align}
    &\frac{16\pi}{\lambda\|\psi\|_g}|A_0\psi|(x,y) ~\le~ 
    1_{(x-y)^2>0, x^0>y^0} \int_{0}^{y^0} d\rho \, \frac{ g(y^0-\rho)}{|\vx-\vy|}\nonumber\\
    &\times \left[ g_2\left(\frac{x^0+y^0 + |\vx-\vy|}{2}\right)
    -g_2\max\left(\frac{x^0+y^0- |\vx-\vy|}{2},\frac{x^0+y^0+ |\vx-\vy|}{2}-\rho\right)\right]\nonumber\\
    &+ 1_{(x-y)^2<0}~1_{ x^0+y^0>|\vx-\vy|} \int_{\frac{y^0-x^0+|\vx-\vy|}{2}}^{y^0} d\rho \, \frac{ g(y^0-\rho)}{|\vx-\vy|}\nonumber\\
    &\times \left[ g_2\max\left(x^0,y^0-\rho\right)
    -g_2\max\left(\frac{x^0+y^0-|\vx-\vy|}{2},\frac{x^0+y^0+|\vx-\vy|}{2}-\rho\right)\right]\nonumber\\
    &+ 1_{(x-y)^2>0, x^0>y^0} \int_{0}^{y^0} d\rho \, \frac{ g(y^0-\rho)}{|\vx-\vy|}\nonumber\\
    &\times \left[ g_2\left(\frac{x^0+y^0 - |\vx-\vy|}{2}-\rho\right)
    -g_2\min\left(\frac{x^0+y^0 - |\vx-\vy|}{2},\frac{x^0+y^0 + |\vx-\vy|}{2}-\rho\right)\right]\nonumber\\\nonumber
    &+ 1_{(x-y)^2<0}~1_{ x^0+y^0>|\vx-\vy|} \int_{\frac{y^0-x^0+|\vx-\vy|}{2}}^{y^0} d\rho \, \frac{ g(y^0-\rho)}{|\vx-\vy|}\nonumber\\
    &\times \left[ g_2\min\left(x^0,y^0-\rho\right)
    -g_2\min\left(\frac{x^0+y^0- |\vx-\vy|}{2},\frac{x^0+y^0+ |\vx-\vy|}{2}-\rho\right)\right]\nonumber\\
    &+1_{(x-y)^2>0,y^0>x^0} \int_{\frac{y^0-x^0-|\vx-\vy|}{2}}^{\frac{x^0+y^0+|\vx-\vy|}{2}}d\rho \, \frac{g(y^0-\rho)}{|\vx-\vy|}\nonumber\\
    &\times \left[g_2 \min\left(\frac{x^0+y^0+|\vx-\vy|}{2}-\rho,x^0\right)-g_2\max\left( \frac{x^0+y^0-|\vx-\vy|}{2}-\rho,0 \right) \right]\nonumber\\
    &+1_{(x-y)^2<0,|\vx-\vy|<x^0+y^0}
    \int_{0}^{y^0}d\rho \,\frac{g(y^0-\rho)}{|\vx-\vy|}\nonumber\\
    &\times \left[g_2 \min\left(\frac{x^0+y^0-|\vx-\vy|}{2},y^0-\rho\right)-g_2\max\left( \frac{x^0+y^0-|\vx-\vy|}{2}-\rho,0\right)  \right].
\end{align}
In order to simplify the result, we introduce the variables
\begin{align}
    \xi^+:=\frac{x^0+y^0+|\vx-\vy|}{2},\\
    \xi^-:=\frac{x^0+y^0-|\vx-\vy|}{2}.
\end{align}
Moreover, we collect terms with the same indicator functions. This results in:
\begin{align}\nonumber
    &\frac{16\pi}{\lambda\|\psi\|_g}|A_0\psi|(x,y)~\le~
    1_{(x-y)^2<0,\xi^->0} \int_0^{y^0}d\rho \, \frac{g(y^0-\rho)}{|\vx-\vy|}
    \Big[g_2\min(\xi^-,y^0-\rho)-g_2\max(\xi^--\rho,0)
    \\\label{A psi estimate 1}
    &+ 1_{\frac{y^0-x^0+|\vx-\vy|}{2}<\rho} \big(g_2(x^0)+g_2(y^0-\rho)
    -g_2(\xi^-)-g_2(\xi^+-\rho)\big)
    \Big]\\\label{A psi estimate 2}
    &+1_{(x-y)^2>0,x^0>y^0}\int_0^{y^0} d\rho \, \frac{g(y^0-\rho)}{|\vx-\vy|} \big[g_2(\xi^+)+g_2(\xi^--\rho)- g_2(\xi^-)-g_2(\xi^+-\rho)\big]\\\label{A psi estimate 3}
    &+1_{(x-y)^2>0,y^0>x^0}\int_{\frac{y^0-x^0-|\vx-\vy|}{2}}^{\xi^+}d\rho \, \frac{g(y^0-\rho)}{|\vx-\vy|}
    \big[g_2\min(\xi^+-\rho,x^0) - g_2\max(\xi^--\rho,0)\big].
\end{align}
This estimate is an important stepping stone in the proof. Except for special weight functions, the resulting expressions are too complicated to be computed explicitly. We therefore continue with further estimates. The main difficulty in these estimates is that the $1/|\vx-\vy|$ singularity in the expressions needs to be compensated by the integrand and that this cancellation needs to be preserved by the respective estimate. Fortunately, the mean value theorem turns out suitable to provide such estimates.

\paragraph{Simplification of \eqref{A psi estimate 1}-\eqref{A psi estimate 3}.}
First, we note that since $g, g_1$ and $g_2$ are monotonously increasing and since $\xi^- \leq \xi^+$, we have in \eqref{A psi estimate 2}:
\be
	g_2(\xi^- - \rho) - g_2(\xi^+ - \rho) \leq 0.
\ee
As the remaining terms in \eqref{A psi estimate 2} still vanish in the limit $|\vx-\vy| \rightarrow 0$, we may replace this difference by zero to obtain a suitable estimate.

Similarly, a brief calculations shows that we have $\xi^+ > y^0$ for $(x-y)^2<0$. It follows that:
\be
	g_2(y^0-\rho) - g_2(\xi^+ - \rho) < 0.
\ee
We shall use this in \eqref{A psi estimate 1}.

Further simplifications can be obtained using the mean value theorem. We begin with the expression in the square brackets in \eqref{A psi estimate 3}. The mean value theorem then implies that there is a \(\chi\in[\max(\xi^--\rho,0),\min(\xi^+-\rho,x^0)]\) such that
\begin{equation}
    g_2\min(\xi^+-\rho,x^0)-g_2\max(\xi^--\rho,0)= \big[\min(\xi^+-\rho,x^0)-\max(\xi^--\rho,0)\big]g_1(\chi).
\end{equation}
Therefore, we have:
\begin{align}
   & g_2\min(\xi^+-\rho,x^0)-g_2\max(\xi^--\rho,0)\nonumber\\
   &~~~\le~ \min(\xi^+-\xi^-,\xi^+-\rho,x^0-\xi^-+\rho,x^0)\, g_1\min(\xi^+-\rho,x^0)\nonumber\\
   & ~~~\le~ |\vx-\vy|\, g_1\min(\xi^+-\rho,x^0) ~\le~ |\vx-\vy| \,g_1(x^0).
\end{align}
Note that the factor $|\vx-\vy|$ exactly compensates the $1/|\vx-\vy|$ singularity. This is the main reason the mean value theorem is so useful here.

Analogously we find for the expression in the square bracket in the first line of \eqref{A psi estimate 1}:
\begin{align}
    &g_2\min(\xi^-,y^0-\rho)-g_2\max(\xi^--\rho,0)\nonumber\\
&~~~\leq ~ \big[\min(\xi^-,y^0-\rho)-\max(\xi^--\rho,0)\big] g_1\min(\xi^-,y^0-\rho)\nonumber\\
    &~~~=~\min(\rho,\xi^-,y^0-\xi^-,y^0-\rho) \, g_1\min(\xi^-,y^0-\rho)\nonumber\\
    &~~~\le~ (y^0-\xi^-) \,g_1\min(\xi^-,y^0-\rho)\nonumber\\
    &~~~\le~ |\vx-\vy| \,g_1\min(\xi^-,y^0-\rho),
\end{align}
where we have used that the further restriction of that term, $(x-y)^2<0$, implies $|\vx-\vy|>|x^0-y^0|\geq y^0-x^0$.

With these considerations, we obtain a rougher but simpler estimate than \eqref{A psi estimate 1}-\eqref{A psi estimate 3}:
\begin{align}\label{massless_after estimate1}
    \frac{16\pi}{\lambda\|\psi\|_g}|A_0\psi|(x,y) ~&\le~
    1_{(x-y)^2<0,\xi^->0} \int_0^{y^0}d\rho~ g(y^0-\rho)\Big[ g_1\min(\xi^-,y^0-\rho) \\\label{massless_after estimate2}
    &~~~+ 1_{\frac{y^0-x^0+|\vx-\vy|}{2}<\rho}\frac{g_2(x^0)-g_2(\xi^-)}{|\vx-\vy|}\Big]\\\label{massless_after estimate3}
    &~~~+1_{(x-y)^2>0,x^0>y^0}\frac{g_2(\xi^+)-g_2(\xi^-)}{|\vx-\vy|} \int_{0}^{y^0}d\rho~ g(y^0-\rho)\\\label{massless_after estimate4}
    &~~~+1_{(x-y)^2>0,y^0>x^0}\, g_1(x^0)\int_{\frac{y^0-x^0-|\vx-\vy|}{2}}^{\xi^+}d\rho~ g(y^0-\rho).
\end{align}
Next, we continue estimating these terms separately so that only expressions without integrals remain.

\paragraph{Further estimate of \eqref{massless_after estimate1}.}
Using the monotonicity of $g_1$ as well as $\min(\xi^-,y^0-\rho) \leq \xi^-$, we find:
\be
	\eqref{massless_after estimate1} ~\leq~1_{(x-y)^2<0,\xi^->0}\, g_1(\xi^-) \int_0^{y^0} ds~ g(s) ~=~  1_{(x-y)^2<0,\xi^->0} \, g_1(\xi^-) g_1(y^0).
\ee
For the constraints given by the indicator function, we have $\xi^- < x^0$. Thus:
\be
	\eqref{massless_after estimate1} \leq 1_{(x-y)^2<0,\xi^->0} \, g_1(x^0) g_1(y^0).
\label{eq:resultmasslessafterestimate1}
\ee

\paragraph{Further estimate of \eqref{massless_after estimate2}.}
We have:
\begin{align}
	 \eqref{massless_after estimate2} ~&=~ 1_{(x-y)^2 < 0, \xi^- >0} \, \frac{ g_2(x^0) - g_2(\xi^-) }{|\vx-\vy|} \int_{\frac{y^0-x^0+|\vx-\vy|}{2}}^{y_0} d \rho \, g(y^0-\rho) \nonumber\\
&= 1_{(x-y)^2 < 0, \xi^- >0} \, \frac{g_2(x^0) - g_2(\xi^-)}{|\vx-\vy|} \int_{0}^{\xi^-} d s \, g(s)\nonumber\\
&= 1_{(x-y)^2 < 0, \xi^- >0} \, \frac{g_2(x^0) - g_2(\xi^-)}{|\vx-\vy|} \big[ g_1(\xi^-) - \underbrace{g_1(0)}_{=0} \big].
\label{eq:secondline159b}
\end{align}
Applying the mean value theorem to $g_2$ in the interval $[\xi^-,x^0]$ (note that here $\xi^-<x^0$), we obtain that:
\be
	\eqref{massless_after estimate2} \leq 1_{(x-y)^2 < 0, \xi^- >0} \, \frac{x^0-\xi^-}{|\vx-\vy|}\, g_1(x^0) g_1(\xi^-).
\ee
Next, we use that $\frac{x^0-\xi^-}{|\vx-\vy|}  = \frac{x^0-y^0+|\vx-\vy|}{2|\vx-\vy|} \leq 1$ as $|x^0-y^0| < |\vx-\vy|$. Thus:
\be
	\eqref{massless_after estimate2} \leq 1_{(x-y)^2 < 0, \xi^- >0} \, g_1(x^0) g_1(\xi^-).
\ee
Using also that for the given constrains $\xi^- < y^0$, we finally obtain:
\be
	\eqref{massless_after estimate2} \leq 1_{(x-y)^2 < 0, \xi^- >0} \, g_1(x^0) g_1(y^0).
\label{eq:resultmasslessafterestimate2}
\ee

\paragraph{Further estimate of \eqref{massless_after estimate3}.}
Here, we can directly carry out the remaining integral using the definition of $g_1$ as the integral of $g$:
\be
	\eqref{massless_after estimate3} = 1_{(x-y)^2>0,x^0>y^0}\, \frac{g_2(\xi^+)-g_2(\xi^-)}{|\vx-\vy|} g_1(y^0).
\ee
Next, we apply the mean value theorem to $g_2$ in the interval $[\xi^-,\xi^+]$ noting that $\xi^+-\xi^- = |\vx-\vy|$. This implies:
\be
	\eqref{massless_after estimate3} \leq 1_{(x-y)^2>0,x^0>y^0} \,g_1(\xi^+) g_1(y^0).
\ee
Next, we note that $(x-y)^2 > 0 \Leftrightarrow |x^0-y^0| > |\vx-\vy|$. Together with $x^0>y^0$, we obtain $x^0>y^0 +|\vx-\vy|$ and therefore:
\be
	\xi^+ = \frac{x^0+y^0 +|\vx-\vy|}{2} \leq x^0.
\ee
Thus, we obtain:
\be
	\eqref{massless_after estimate3} \leq 1_{(x-y)^2>0,x^0>y^0} \, g_1(x^0) g_1(y^0).
	\label{eq:resultmasslessafterestimate3}
\ee

\paragraph{Further estimate of \eqref{massless_after estimate4}.}
Here, we carry out the remaining integral as well.
\begin{align}
	\eqref{A psi estimate 3} ~&\leq~ 1_{(x-y)^2>0,y^0>x^0}\, g_1(x^0) [ g_1(\xi^+)-g_1( (y^0-x^0-|\vx-\vy|)/2)]\nonumber\\
&\leq~ 1_{(x-y)^2>0,y^0>x^0}\, g_1(x^0)g_1(y^0).
\label{eq:resultmasslessafterestimate4}
\end{align}
as $\xi^+\leq y^0$.

\paragraph{Summary of the result.} Gathering the terms \eqref{eq:resultmasslessafterestimate1}, \eqref{eq:resultmasslessafterestimate2}, \eqref{eq:resultmasslessafterestimate3} and \eqref{eq:resultmasslessafterestimate4} yields:
\be
	  \frac{16\pi}{\lambda\|\psi\|_g}\, |A_0 \psi |(x,y) \leq g_1(x^0) g_1(y^0) \left( 2 \times 1_{(x-y)^2 < 0, \xi^- >0} + 1_{(x-y)^2>0,x^0>y^0} + 1_{(x-y)^2>0,y^0>x^0}\right).
\ee
Considering that the conditions in different indicator functions are mutually exclusive, we finally obtain:
\be
	  \frac{16\pi}{\lambda\|\psi\|_g}\, |A_0 \psi |(x,y) \leq 2 g_1(x^0) g_1(y^0).
	\label{eq:resultmasslessestimate}
\ee
Dividing by $g(x^0)g(y^0)$, taking the supremum over $x,y \in \tfrac{1}{2}\M$ and factorizing into one-dimensional suprema finally yields the claim \eqref{eq:estimatea0}.

\subsubsection{Estimate of the mixed terms \eqref{eq:estimatea1} and \eqref{eq:estimatea2}.} \label{sec:estimatemixed}

We focus on $A_2$ first, starting from its definition \eqref{eq:defa2}. We take the absolute value and make use of $|\psi(x,y)| \leq g(x^0) g(y^0)\, \| \psi \|_g$. Moreover, we use:
\be
	\left| J_1(t)/t \right| \leq \frac{1}{2}.
\ee
This yields:
\begin{align}
	|A_2 \psi|(x,y) &\leq  \frac{\lambda \, m_2^2 \, \|\psi\|_g}{4(4 \pi)^3} \int d^3 \vx' \int d^3 \vy'~\frac{H(x^0-|\vx-\vx'|)}{|\vx-\vx'|} \frac{g( x^0-|\vx-\vx'|)}{|\vx'-\vy'|} \nonumber\\
&\times \left[ H(x^0 - |\vx-\vx'| + |\vx'-\vy'|)H(y^0-x^0 + |\vx-\vx'| - |\vx'-\vy'| -|\vy-\vy'|) \right. \nonumber\\
&\times g(x^0 - |\vx-\vx'| + |\vx'-\vy'| )\nonumber\\
&+ H(x^0 - |\vx-\vx'| - |\vx'-\vy'|)H(y^0-x^0 + |\vx-\vx'| + |\vx'-\vy'|-|\vy-\vy'|) \nonumber\\
 &\left. \times g(x^0 - |\vx-\vx'| - |\vx'-\vy'|)\right].
\end{align}
As the remaining singularities are independent of each other for a suitable choice of integration variables (see below), we are left with an integrable function on a finite domain.

The next task is to bring the expressions into a simpler form. One possibility to do this is to use
\be
	H(y^0-x^0+|\vx-\vx'|+|\vx'-\vy'| - |\vy-\vy'|) \leq H(y^0-x^0+|\vx-\vx'|+|\vx'-\vy'|)
\ee
for the second Heaviside function in the second summand.
The first Heaviside function in the first summand equals 1 anyway, as $|\vx-\vx'| < x^0$. We furthermore use
\be
	H(y^0-x^0+|\vx-\vx'|-|\vx'-\vy'| - |\vy-\vy'|) \leq H(y^0-x^0+|\vx-\vx'|-|\vx'-\vy'|),
\ee
as it simplifies the domain of integration. Overall, the domain of integration remains bounded. 
Introducing $\vz_1 = \vx-\vx'$, $\vz_2 = \vx'-\vy'$ (with Jacobi determinant of modulus 1) and using spherical coordinates for $\vz_2$, this leads to:
\begin{align}
	&|A_2 \psi|(x,y) \, \frac{4(4\pi)^3}{\lambda \, m_2^2 \, \| \psi \|_g}\nonumber\\
& \leq~ \int_{B_{x^0}(0)} d^3 \vz_1 4\pi \int_0^{\max(0,y^0-x^0+|\vz_1|)}\!\!\!\!\!\! d^3 \vz_2 ~|\vz_2|^2\, \frac{1}{|\vz_1|} \frac{1}{|\vz_2|} g(x^0-|\vz_1|) g(x^0-|\vz_1|+|\vz_2|) \nonumber\\
&+~\int_{B_{x^0}(0)} d^3 \vz_1 ~4\pi \int_{\max( 0, x^0-y^0-|\vz_1|)}^{x^0-|\vz_1|} \!\!\!\!\!\! d|\vz_2|~|\vz_2|^2 ~\frac{1}{|\vz_1|} \frac{1}{ |\vz_2|} g(x^0-|\vz_1|) g(x^0-|\vz_1|-|\vz_2|).
\end{align}

Using spherical coordinates also for $\vz_1$, this can be further simplified to:
\begin{align}
	|A_2 \psi|(x,y) \, &\frac{16\pi}{\lambda \, m_2^2 \, \| \psi \|_g} \leq  \int_0^{x^0} d r_1 \int_0^{\max(0,y^0-x^0+r_1)} \!\!\! dr_2~r_1 r_2\,g(x^0-r_1) g(x^0-r_1+r_2)\label{eq:A2firstestimate}\\
&+~\int_0^{x^0} d r_1 \int_{\max( 0,x^0-r_1-y^0)}^{t_1-r_1} \!\!\! dr_2~r_1 r_2 ~ g(x^0-r_1) g(x^0-r_1-r_2) \label{eq:A2secondestimate}.
\end{align}

Our next task is to simplify the remaining integrals. We begin with making the change of variables $\rho = x^0-r_1$:
\begin{align}
	|A_2 \psi|(x,y) \, \frac{16\pi}{\lambda \, m_2^2 \, \| \psi \|_g} ~\leq~  &\int_0^{x^0} d \rho~(x^0-\rho) g(\rho) \int_0^{\max(0,y^0-\rho)} \!\!\! dr_2~r_2\, g(\rho+r_2)\nonumber\\
&+~\int_0^{x^0} d \rho~(x^0-\rho)g(\rho) \int_{\max( 0,\rho-y^0)}^{\rho} \!\!\! dr_2~r_2 \, g(\rho-r_2).
\label{eq:a2estimate1}
\end{align}
Now we consider the $r_2$-integral in both terms and integrate by parts. This yields:
\begin{align}
	 \int_0^{\max(0,y^0-\rho)} \!\!\! dr_2~r_2\, g(\rho+r_2)&=\max(0,y^0-\rho) g_1(y^0) - g_2(\max(\rho,y^0)) + g_2(\rho),\\
\int_{\max( 0,\rho-y^0)}^{\rho} \!\!\! dr_2~r_2 \, g(\rho-r_2) ~&=~ \max( 0,\rho-y^0) g_1(y^0) + g_2(\min(\rho,y^0)).
\end{align}
We now use $- g_2(\max(\rho,y^0)) + g_2(\rho) \leq 0$ in the first term and then re-insert the resulting estimate into \eqref{eq:a2estimate1}. Considering also $\max(0,y^0-\rho) + \max( 0,\rho-y^0) = |y^0-\rho|$, this yields:
\be
	|A_2 \psi|(x,y) \, \frac{16\pi}{\lambda \, m_2^2 \, \| \psi \|_g} ~\leq~  \int_0^{x^0} d \rho~(x^0-\rho) g(\rho) \left[|y^0-\rho| g_1(y^0) + g_2(\min(\rho,y^0)) \right]
\label{eq:a2estimate2}
\ee
The first summand of \eqref{eq:a2estimate2} can be treated as follows. First we focus on whether $x^0>y^0$ or $x^0\leq y^0$. In the first case, we then differentiate between the cases $\rho < y^0$ and $\rho \geq y^0$ and split up the integrals accordingly. This yields:
\begin{align}
	 &\int_0^{x^0} d \rho~(x^0-\rho) g(\rho) |y^0-\rho| g_1(y^0)\nonumber\\
&=~ g_1(y^0)\, H(x^0-y^0) \int_0^{y^0} d \rho~(x^0-\rho)(y^0-\rho) g(\rho)\label{eq:a2estimate2a}\\
&~~~ - g_1(y^0)\, H(x^0-y^0) \int_{y^0}^{x^0} d \rho~(x^0-\rho)(y^0-\rho) g(\rho)\label{eq:a2estimate2b}\\
& ~~~+ g_1(y^0)\, H(y^0-x^0) \int_0^{x^0} d \rho~(x^0-\rho)(y^0-\rho) g(\rho).
\label{eq:a2estimate2c}
\end{align}
We now calculate these terms separately using integration by parts. The first term yields:
\begin{align}
	\eqref{eq:a2estimate2a} ~&=~ g_1(y^0) H(x^0-y^0) \left[ (x^0-y^0) g_2(y^0)+ 2g_3(y^0)\right].
\label{eq:a2estimate2a2}
\end{align}
We turn to \eqref{eq:a2estimate2b}:
\begin{align}
	\eqref{eq:a2estimate2b} ~&=~ -g_1(y^0) H(x^0-y^0)\left[ (y^0-x^0)(g_2(x^0)+g_2(y^0))+2g_3(x^0) - 2g_3(y^0)\right].
\label{eq:a2estimate2b2}
\end{align}
The result of \eqref{eq:a2estimate2c} is:
\begin{align}
	\eqref{eq:a2estimate2c} ~&=~ g_1(y^0) H(y^0-x^0) \left[ (y^0-x^0) g_2(x^0)+ 2g_3(x^0)\right].
	\label{eq:a2estimate2c2}
\end{align}
Gathering the terms \eqref{eq:a2estimate2a2}, \eqref{eq:a2estimate2b2} and \eqref{eq:a2estimate2c2} yields:
\begin{align}
	|A_2 \psi|(x,y) \, \frac{16\pi}{\lambda \, m_2^2 \, \| \psi \|_g} ~&\leq~ g_1(y^0) H(x^0-y^0) \left[ 2(x^0-y^0)g_2(y^0) + 4 g_3(y^0)-2g_3(x^0) \right]\nonumber\\
&~~~+g_1(y^0) |x^0-y^0| g_2(x^0) + 2 g_1(y^0) H(y^0-x^0) g_3(x^0)\nonumber\\
&\leq~ 2g_1(y^0)|x^0-y^0| g_2(x^0) + 2 g_1(y^0)g_3(x^0) H(x^0-y^0)\nonumber\\
&~~~+g_1(y^0)|x^0-y^0|g_2(x^0) + 2 g_1(y^0) g_3(x^0) H(y^0-x^0)\nonumber\\
&=~ 3g_1(y^0)|x^0-y^0| g_2(x^0) + 2 g_1(y^0)g_3(x^0)\nonumber\\
&\leq~3(x^0+y^0)g_1(y^0) g_2(x^0) + 2 g_1(y^0)g_3(x^0).
\end{align}
In order to obtain $\| A_2 \psi \|_g$, we divide by $g(x^0)g(y^0)$ and take the supremum over $x,y \in \tfrac{1}{2}\M$. This results in:
\begin{align}
	\sup_{\psi \in \mathcal{S}((\frac{1}{2}\M)^2)} \frac{\| A_2 \psi \|_g}{\| \psi \|_g} ~&\leq~ \frac{\lambda \, m_2^2}{16\pi} \left( 3\sup_{x^0,y^0 \geq 0} \frac{(x^0+y^0)g_2(x^0)\, g_1(y^0)}{g(x^0)g(y^0)}  + 2 \sup_{x^0,y^0\geq 0} \frac{g_3(x^0)g_1(y^0)}{g(x^0)g(y^0)}\right).
\end{align}
After factorizing the two-dimensional suprema into one-dimensional ones, this exactly yields the claim, \eqref{eq:estimatea2}.

For the operator $A_1$, we find analogously:
\begin{align}
	\sup_{\psi \in \mathcal{S}((\frac{1}{2}\M)^2)} \frac{\| A_1 \psi \|_g}{\| \psi \|_g} ~&\leq~ \frac{\lambda \, m_1^2}{16\pi} \left( 3\sup_{x^0,y^0 \geq 0} \frac{(x^0+y^0)g_1(x^0)\, g_2(y^0)}{g(x^0)g(y^0)}  + 2 \sup_{x^0,y^0\geq 0} \frac{g_1(x^0)g_3(y^0)}{g(x^0)g(y^0)}\right).
\end{align}
which, after factorization into one-dimensional suprema, yields the claim \eqref{eq:estimatea1}.

\subsubsection{Estimate of the mass-mass term \eqref{eq:estimatea12}.} \label{sec:estimatemassmass}

We begin with \eqref{eq:defa12}. Taking the absolute value and using $|\psi(x,y)| \leq \| \psi \|_g \, g(x^0) g(y^0)$ as well as $|J_1(t)/t|\leq \frac{1}{2}$ yields:
\begin{align}
&|A_{12} \psi|(x,y) \leq \frac{\lambda \, m_1 m_2 \, \| \psi \|_g}{4(4\pi)^3}  \int_0^\infty d{x'}^0 \int d^3 \vx' \int_0^\infty d{y'}^0 \int_0^{2\pi} d\varphi \int_{0}^{\pi} d \vartheta \, \cos(\vartheta) |{x'}^0-{y'}^0| \,  \nonumber\\
&\times H(x^0-{x'}^0-|\vx-\vx'|)H(y^0-{y'}^0-|\vy-\vx'+\vz|)g({x'}^0)g({y'}^0)\Big|_{|\vz| = |{x^0}'-{y^0}'|},
\end{align}
where, we recall, $\vz$ is the variable for which the spherical coordinates are used.

Next, we consider the ranges of integration which the Heaviside functions imply. $H(x^0-{x'}^0-|\vx-\vx'|)$ restricts the range of integration of $\vx'$ to the ball $B_{x^0-{x'}^0}(\vx)$ and the range of the ${x'}^0$-integration to $(0,x^0)$. The range implied by the second Heaviside function is more complicated. We therefore use the estimate
\be
	H(y^0-{y'}^0-|\vy-\vx'+\vz|) \leq H(y^0-{y'}^0).
\ee
Then ${y'}^0 \in (0,y^0)$ and there is no further restriction for the angular variables. We obtain:
\begin{align}
|A_{12} \psi|(x,y) &\leq \frac{\lambda \, m_1 m_2 \, \| \psi \|_g}{8(4\pi)^3}  \int_0^{x^0} d{'}^0 \int_{B_{x^0-{x'}^0}(\vx)} \!\!\!\!\!\!\!\!\!\! d^3 \vx' \int_0^{y^0} d{y'}^0 \int_0^{2\pi} d\varphi \int_{0}^{\pi} d \vartheta \nonumber\\
&~~~\times \cos(\vartheta) |{x'}^0-{y'}^0| \, g({x'}^0)g({y'}^0).
\end{align}
Performing the $\vx'$-integration, as well as the angular integrals yields:
\be
	|A_{12} \psi|(x,y) \leq \frac{\lambda \, m_1 m_2 \, \| \psi \|_g}{96\pi}  \int_0^{x^0} d{x'}^0 |x^0-{x'}^0|^3 g({x'}^0) \int_0^{y^0} d{y'}^0~ |{x'}^0-{y'}^0| \, g({y'}^0).
\ee
Our next task is to estimate the term explicitly in terms of the functions $g_n$ only. To do so, we use
\be
	|{x'}^0-{y'}^0| \leq {x'}^0 + {y'}^0.
\ee
This yields:
\be
	|A_{12} \psi|(x,y) \leq \frac{\lambda \, m_1 m_2 \, \| \psi \|_g}{48\pi}  \int_0^{x^0} d{x'}^0 |x^0-{x'}^0|^3 g({x'}^0) \int_0^{y^0} d{y'}^0~ ({x'}^0 + {y'}^0) g({y'}^0).
\label{eq:a12estimatecalc}
\ee
Let
\be
	I(x^0,y^0) = \int_0^{x^0} d{x'}^0 |x^0-{x'}^0|^3 g({x'}^0) \int_0^{y^0} d{y'}^0~ ({x'}^0 + {y'}^0) g({y'}^0)
\ee
and
\be
	L({x'}^0,y^0) = \int_0^{y^0} d{y'}^0~ ({x'}^0 + {y'}^0) g({y'}^0).
\ee
Integration by parts yields:
\be
	L({x'}^0,y^0) = {x'}^0 g_1(y^0) + y^0 g_1(y^0) - g_2(y^0) \leq {x'}^0 g_1(y^0) + y^0 g_1(y^0).
\ee
Next, let
\begin{align}
	I_a(x^0) &= \int_0^{x^0} d {x'}^0~|x^0-{x'}^0|^3 g({x'}^0),\nonumber\\
	I_b(x^0) &=  \int_0^{x^0} d {x'}^0~{x'}^0 |x^0-{x'}^0|^3 g({x'}^0).
\end{align}
Then:
\be
	I(x^0,y^0) \leq I_a(x^0) \, y^0 g_1(y^0) + I_b(x^0) \, g_1(y^0).
\label{eq:massmassintparts}
\ee
We consider $I_a$ first, using $(x^0-{x'}^0)^2 \leq (x^0)^2$ and integrating by parts:
\begin{align}
	I_a(x^0) &\leq (x^0)^2 \int_0^{x^0} d {x'}^0~(x^0-{x'}^0)g({x'}^0)\nonumber\\
&= (x^0)^2 \left( \underbrace{(x^0-{x'}^0)g_1({x'}^0)|_{{x'}^0 = 0}^{x^0}}_{=0} + g_2(x^0)\right) = (x^0)^2 g_2(x^0).
\end{align}
We turn to $I_b$, using ${x'}^0(x^0-{x'}^0)\leq \frac{1}{4}(x^0)^2$ and integrating by parts twice. This results in:
\begin{align}
	I_b(x^0) ~\leq~ \frac{(x^0)^2}{4} \int_0^{x^0} d{x'}^0~ (x^0-{x'}^0)^2 g({x'}^0) ~=~ \frac{(x^0)^2}{2}\, g_3(x^0).
\end{align}
Considering \eqref{eq:massmassintparts}, we therefore obtain:
\be
	I(x^0,y^0) ~\leq~ (x^0)^2 g_2(x^0)\, y^0 g_1(y^0) + \frac{(x^0)^2}{2}\, g_3(x^0)\, g_1(y^0).
\label{eq:resultmassmasstermgeneralg}
\ee
Returning to \eqref{eq:a12estimatecalc}, we divide by $g(x^0)g(y^0)$ and take the supremum, with the result:
\begin{align}
	\sup_{\psi \in \mathcal{S}((\frac{1}{2}\M)^2)} \frac{\| A_{12} \psi \|_g}{\| \psi \|_g} ~&\leq~\frac{\lambda \, m_1 m_2 \, \| \psi \|_g}{96\pi} \left[ \sup_{x^0,y^0\geq 0} \frac{(x^0)^2 g_2(x^0)\, y^0 g_1(y^0)}{g(x^0)g(y^0)} \right.\nonumber\\
&~~~\left.+ \frac{1}{2} \sup_{x^0,y^0\geq 0} \frac{(x^0)^2 g_3(x^0)\, g_1(y^0)}{g(x^0)g(y^0)} \right].
\end{align}
Factorizing the two-dimensional suprema into one-dimensional ones yields the claim, \eqref{eq:estimatea12}.

\subsection{Proof of Theorem \ref{thm:exponentialg}} \label{sec:proofexponentialg}

Let $\psi \in \mathcal{S}$. It only remains to calculate the supremum in \eqref{eq:estimatea0} for $g(t)=e^{\gamma t}$. We have:
\be
	g_1(t) = \frac{1}{\gamma} \left( e^{\gamma t} - 1\right)
\ee
and hence
\be
	\sup_{\psi \in \mathcal{S}((\frac{1}{2}\M)^2)} \frac{\| A_0 \psi \|_g}{\| \psi \|_g} ~\leq~ \frac{\lambda}{8\pi} \left( \sup_{t \geq 0} \frac{g_1(t)}{g(t)} \right)^2 ~=~ \frac{\lambda}{4\pi} \left( \sup_{t \geq 0} \frac{1}{\gamma} (1 - e^{-\gamma t}) \right)^2~=~ \frac{\lambda}{8\pi \gamma^2}. 
\ee
This shows that $A_0$ can be linearly extended to a bounded operator on $\Banach_g$ which satisfies the same estimate, \eqref{eq:norma0exponential}. Moreover, for $\gamma > \sqrt{\frac{\lambda}{4\pi}}$, $A_0$ is a contraction and Banach's fixed point theorem implies the existence of a unique solution $\psi \in \Banach_g$ of the equation $\psi = \psi^\free + A_0 \psi$ for every $\psi^\free \in \Banach_g$.

\subsection{Proof of Theorem \ref{thm:existence}} \label{sec:proofexistence}

Let again $\psi \in \mathcal{S}$. We need to calculate the suprema in \eqref{eq:estimatea0}-\eqref{eq:estimatea12} for $g(t)=(1+\alpha t^2)e^{\alpha t^2/2}$. We first note:
\begin{align}
	g_1(t) ~&=~ t e^{\alpha t^2/2},\nonumber\\
	g_2(t) ~&=~\frac{1}{\alpha} \left( e^{\alpha t^2/2}-1 \right),\nonumber\\
	g_3(t) ~&=~\frac{1}{\alpha} \left[ \sqrt{\frac{\pi}{2\alpha}} \erfi(\sqrt{\alpha/2} t)-t \right].
\end{align}
We can see that with each successive integration, the functions $g_n$ grow slower as $t\rightarrow \infty$. Furthermore, the leading terms in $g_n$ are inversely proportional to increasing powers of $\alpha$. These two properties (and of course the fact that $g_1, g_2,g_3$ can be written down in terms of elementary functions) make this particular function $g(t)$ a suitable choice for the proof.

As we need to estimate the behavior of quotients like $g_3(t)/g(t)$ for $t\rightarrow \infty$, we look for a simpler estimate of $g_3$ in terms of exponential functions. We note:
\begin{align}
	g_3(t) ~&=~ \int_0^t dt'\,  \frac{1}{\alpha} \left( e^{\alpha {t'}^2/2} -1\right)\nonumber\\
	&\le~ \frac{e^{\alpha t^2/2}}{\alpha} e^{-\alpha t^2/2} \sqrt{2/\alpha} \int_0^{\sqrt{\alpha/2} t}  d\tau  \,e^{\tau^2}\nonumber\\
	&=~\frac{\sqrt{2}}{\alpha^{3/2}} \, e^{\alpha t^2/2} \, D(\sqrt{\alpha/2}\,t),
\end{align}
where $D(t) = e^{-t^2}\int_0^{t}  d\tau \, e^{\tau^2}$ denotes the Dawson function.
Using the property \(|tD(t)|<\frac{2}{3}\), we obtain:
\be
	t g_3(t)~\leq~\frac{4}{3} \frac{e^{\alpha t^2/2}}{\alpha^{2}}.
\label{eq:g3estimate}
\ee
We are now well-equipped to calculate the suprema occurring in \eqref{eq:estimatea0}-\eqref{eq:estimatea12}. Using
 \begin{equation}
 \sup_{t\geq 0}\frac{t^\beta}{1+t^2} ~=~ \left\{ \begin{matrix} 1\quad \text{ for } \beta=0\\ \frac{1}{2} \quad \text{ for } \beta=1 \\ 1\quad \text{ for } \beta=2 \end{matrix}\right. 
 \end{equation}
we obtain:
\begin{align}
	\sup_{t\geq 0} \frac{g_1(t)}{g(t)} ~&=~ \sup_{t\geq 0} \frac{t}{1+\alpha t^2} ~=~ \frac{1}{2}\frac{1}{\sqrt{\alpha}},\label{eq:sup1}\\
	\sup_{t\geq 0} \frac{t g_1(t)}{g(t)} ~&=~ \sup_{t\geq 0} \frac{t^2}{1+\alpha t^2} ~=~ \frac{1}{\alpha},\label{eq:sup2}\\
	\sup_{t\geq 0} \frac{g_2(t)}{g(t)} ~&\leq~ \sup_{t\geq 0} \frac{1}{\alpha} \frac{1}{1+\alpha t^2} ~=~ \frac{1}{\alpha},\label{eq:sup3}\\
\sup_{t\geq 0} \frac{tg_2(t)}{g(t)} ~&\leq~ \sup_{t\geq 0} \frac{1}{\alpha} \frac{t}{1+\alpha t^2} ~=~ \frac{1}{2}\frac{1}{\alpha^{3/2}},\label{eq:sup4}\\
\sup_{t\geq 0} \frac{t^2g_2(t)}{g(t)} ~&\leq~ \sup_{t\geq 0} \frac{1}{\alpha} \frac{t^2}{1+\alpha t^2} ~=~ \frac{1}{\alpha^2}\label{eq:sup5}.
\end{align}
Using, in addition, the property $|D(t)| < \frac{3}{5}$, we find:
\begin{align}
	\sup_{t\geq 0} \frac{g_3(t)}{g(t)} ~&\leq~ \sup_{t\geq 0} \frac{\sqrt{2}}{\alpha^{3/2}} \frac{D(\sqrt{\alpha/2}t)}{1+\alpha t^2} ~=~ \frac{3\sqrt{2}}{5}\frac{1}{\alpha^{3/2}} ~<~ \frac{1}{\alpha^{3/2}},\label{eq:sup6}\\
\sup_{t\geq 0} \frac{t^2 g_3(t)}{g(t)} ~&\leq~ \sup_{t\geq 0} \frac{4}{3}\frac{1}{\alpha^2} \frac{t}{1+\alpha t^2} ~=~\frac{2}{3}\frac{1}{\alpha^{5/2}}. \label{eq:sup7}
\end{align}
In the last line, we have made use of \eqref{eq:g3estimate}.

With these results, we find for $A_0$:
\be
	\eqref{eq:estimatea0} ~\leq~ \frac{\lambda}{8\pi} \left(\frac{1}{2} \frac{1}{\sqrt{\alpha}}\right)^2 ~=~  \frac{\lambda}{32\pi} \frac{1}{\alpha}.
\ee
This yields \eqref{eq:estimatea0final}.

We continue with $A_1$.
\be
	\eqref{eq:estimatea1} ~\leq~ \frac{\lambda \, m_1^2}{16\pi} \left[  3 \, \frac{1}{\alpha} \, \frac{1}{\alpha} + 3 \, \frac{1}{2} \frac{1}{\sqrt{\alpha}} \, \frac{1}{2} \frac{1}{\alpha^{3/2}} +2 \, \frac{1}{2} \frac{1}{\sqrt{\alpha}} \, \frac{1}{\alpha^{3/2}} \right] ~=~ \frac{\lambda \, m_1^2}{16\pi}\, \frac{19}{4} \frac{1}{\alpha^2} ~<~\frac{\lambda \, m_1^2}{16\pi} \frac{5}{\alpha^2}.
\ee
This yields \eqref{eq:estimatea1final}. Analogously, we obtain the estimate \eqref{eq:estimatea2final} for $A_2$.

Finally, for $A_{12}$, we have
\be
	\eqref{eq:estimatea12} ~\leq~ \frac{\lambda \, m_1^2\, m_2^2}{96 \pi} \left[ \frac{1}{\alpha^2} \, \frac{1}{\alpha} + \frac{1}{2} \, \frac{2}{3} \frac{1}{\alpha^{5/2}} \, \frac{1}{2} \frac{1}{\sqrt{\alpha}} \right] ~=~ \frac{\lambda \, m_1^2\, m_2^2}{96\pi} \, \frac{7}{6} \frac{1}{\alpha^3} ~<~ \frac{\lambda \, m_1^2\, m_2^2}{80 \pi} \, \frac{1}{\alpha^3},
\ee
which yields \eqref{eq:estimatea12final}.

Now, the estimates \eqref{eq:estimatea0final}-\eqref{eq:estimatea12final} show that the operators $A_0$, $A_1$, $A_2$ and $A_{12}$ are bounded on test functions. Thus, they can be linearly extended to bounded operators on $\Banach_g$ with the same bounds.

The operator $A = A_0 + A_1 + A_2 + A_{12}$ then also defines a bounded linear operator on $\Banach_g$ with norm
\be
	\| A \| ~\leq~ \| A_0 \| + \| A_1 \| + \| A_2 \| + \| A_{12} \|.
\ee
Using the previous results \eqref{eq:estimatea0final}-\eqref{eq:estimatea12final}, we obtain:
\be
	\|A \| ~\leq~ \frac{\lambda}{8\pi \alpha} \left( \frac{1}{4} + \frac{5(m_1^2 + m_2^2)}{2} \frac{1}{\alpha} + \frac{m_1^2 \, m_2^2}{10} \frac{1}{\alpha^2} \right).
\ee
If $\alpha$ is chosen such that this expression is strictly smaller than unity, $A$ becomes a contraction and the existence and uniqueness of solutions of the equation $\psi = \psi^\free + A\psi$ follows. This yields condition \eqref{eq:condexistencemassive} and ends the proof.

\subsection{Proof of Theorem \ref{thm:existencecurved}} \label{sec:proofexistencecurved}

The proof can be reduced to the one for $\tfrac{1}{2}\M$. To do so, we take the absolute value of \eqref{eq:defa0tilde} and use $|\psi|(\eta_1,\vx,\eta_2,\vy) \leq g(\eta_1) g(\eta_2) \|\psi\|_g$. With
\begin{align}
	G(\eta) ~&=~ a(\eta) \exp \left(\gamma \int_0^\eta d\eta'~a(\eta') \right)	\label{eq:defG}\\
	G_1(\eta)~&=~ \int_0^\eta d\eta'~G(\eta)
\end{align}
we obtain the estimate
\begin{align}
    &|\widetilde{A}_0\psi|(x,y) \le \frac{\lambda \|\psi\|_g}{4(4\pi)^3} \int_{B_{\eta_2}(\vy)}d^3 \vy' \int_0^{2\pi} d\varphi \int_{-1}^1 d\cos\vartheta \, \frac{|b^2|}{(b^0+|\vb|\cos\vartheta)^2 |\vy'|} G(\eta_2-|\vy'|)\nonumber\\
    &\times G\left(\eta_1-\frac{1}{2}\frac{b^2}{b^2+|\vb|\cos\vartheta}\right)
    \left(1_{b^2>0}1_{b^0>0} 1_{\cos\vartheta > \frac{b^2}{2\eta_1^0|\vb|} - \frac{b^0}{|\vb|}}+1_{b^2<0}1_{\cos\vartheta<\frac{b^2}{2\eta_1|\vb|} - \frac{b^0}{|\vb|}}\right).
\label{eq:a0tildecalc01}
\end{align}
This estimate is identical to \eqref{eq:a0calc01} with the only difference that the function $g$ is exchanged with $G$ in the integral (but not in $\| \cdot\|_g$). Thus, going through the same steps as in Secs. \ref{sec:proofbounds}, \ref{sec:proofexistence}, we obtain:
\be
	\sup_{\psi \in \mathcal{S}\left(([0,\infty)\times \R^3)^2\right)} \frac{\| \widetilde{A}_0 \psi \|_g}{\| \psi \|_g} ~\leq~ \frac{\lambda}{8\pi} \left(\sup_{t\geq 0} \frac{G_1(t)}{g(t)}\right)^2.\label{eq:estimatea0tilde}
\ee
Now, recalling $g(t) = \exp\left(\gamma \int_0^t d\tau \, a(\tau)\right)$ we have
\be
	G_1(t) = \frac{1}{\gamma} g(t)
\ee
and it follows that
\be
	\sup_{\psi \in \mathcal{S}\left(([0,\infty)\times \R^3)^2\right)} \frac{\| \widetilde{A}_0 \psi \|_g}{\| \psi \|_g} ~\leq~ \frac{\lambda}{8\pi \gamma^2},
\ee
which yields \eqref{eq:a0tildebound}. The rest of the claim follows as before.

\section{Conclusions}
\label{sec:conclusions}

In this paper we have given what we think of as a satisfactory answer to the problem posed: to prove the existence and uniqueness of solution of the integral equation \eqref{eq:inteq} and its $N$-particle generalization \eqref{eq:npartint}. Following previous works, we have assumed a cutoff in time. By considering an example for our integral equation on a cosmological spacetime with a Big Bang singularity, we have shown that such a cutoff can arise naturally and without violating any spacetime symmetries.

Our work provides a rigorous proof of the existence of interacting relativistic quantum dynamics in 1+3 spacetime dimensions; in particular, our model does not suffer from ultraviolet divergences which are typically encountered in quantum field theoretic models. Of course, our model does not describe particle creation and annihilation and is therefore a toy model rather than an alternative to QFT. Nevertheless, we find the fact that direct interactions, even singular ones along the light cone, can be made mathematically rigorous, remarkable. We wonder whether in the long run the mechanism of interaction through multi-time integral equations and direct interactions could contribute to a rigorous formulation of quantum field theory.

In the more immediate future, it would first of all be desirable to extend our results to Dirac particles (meaning that the Green's functions in \eqref{eq:inteq} are replaced with Green's functions of the Dirac equation). As the Dirac Green's functions involve distributional derivatives, it would then be more difficult than in the Klein-Gordon case to define the combination of the three distributions $G_1^\ret$, $G_2^\ret$ and $\delta((x-y)^2)$ which occur in \eqref{eq:inteq}. Moreover, as the previous work \cite{int_eq_dirac} on Dirac particles but regular interaction kernels $K$ suggests, the occurrence of the distributional derivatives in the Green's functions alone leads to technical complications, as one then needs to prove a higher regularity of the solutions. In the $N$-particle case, this regularity would have to be greater than in the two-particle case so that one cannot simply add up estimates for the norm of the two-particle integral operator to obtain an estimate for the $N$-particle integral operator anymore. There would be further terms to consider.
This is the set of questions which a work on the Dirac case of Eq.\@ \eqref{eq:inteq} would have to answer.

Besides the Dirac case, there is also a range of more detailed technical questions for the Klein-Gordon case which would be interesting to address. While we have here worked with a weighted $L^\infty$ norm both for time and space variables, one could also try to use a weighted $L^\infty L^2$ norm instead ($L^\infty$ for the time variables and $L^2$ for the space variables). It would then be a challenging task to find the right inequalities to obtain similar estimates as we did. Moreover, one could also try to prove higher regularity not only in the sense of integrability but also differentiability. An interesting question, for example, is whether one can apply the Klein-Gordon operators $(\Box_k + m_k^2)$ to the solutions of \eqref{eq:inteq} in a weak sense. For the Dirac case, an analogous property was, in fact, established in \cite{int_eq_dirac}. 

\subsubsection*{Acknowledgments}
We would like to thank Dirk Deckert and Roderich Tumulka for helpful discussions. M. N. acknowledges funding from Cusanuswerk and from the Elite Network of Bavaria, through the Junior Research Group `Interaction Between Light and Matter'.
M. N. acknowledges funding from Cusanuswerk and from the Elite Network of Bavaria, through the Junior Research Group `Interaction Between Light and Matter'.
\\[1mm]
\begin{minipage}{15mm}
\includegraphics[width=13mm]{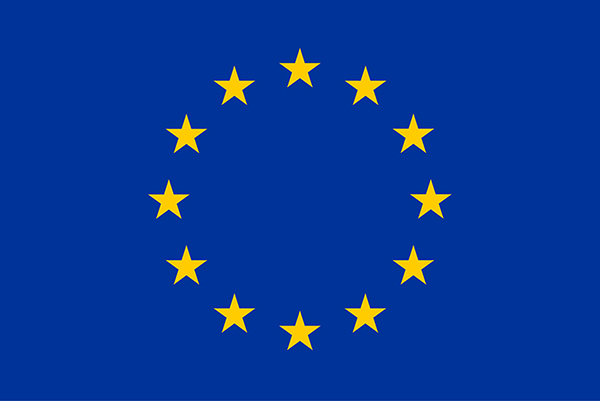}
\end{minipage}
\begin{minipage}{143mm}
This project has received funding from the European Union's Framework for
Re- \\ search and Innovation Horizon 2020 (2014--2020) under the Marie Sk{\l}odowska-
\end{minipage}\\[1mm]
Curie Grant Agreement No.~705295.


\end{document}